%% file: uv_paper_arxiv.tex
\documentclass[
	aps, prd, a4paper, 10pt, notitlepage, 
	amsmath, amssymb, amsfonts, eqsecnum,
	superscriptaddress, 
	nofootinbib, longbibliography 
]{revtex4-1}

\usepackage{xcolor}
\usepackage{graphicx}
\usepackage{dsfont}
\usepackage{upgreek}
\usepackage{float}
\usepackage[normalem]{ulem}

\xdefinecolor{mylinkcolor}{rgb}{0 0 0.5}

\usepackage{tikz}
\usetikzlibrary{snakes}
\usetikzlibrary{shapes,arrows}
\usetikzlibrary{shadows}
\usetikzlibrary{positioning}

\usepackage[pdftex,breaklinks=true,colorlinks=true,filecolor=mylinkcolor,citecolor=mylinkcolor,linkcolor=mylinkcolor,urlcolor=mylinkcolor,menucolor=mylinkcolor,bookmarksnumbered,bookmarksopen,bookmarksopenlevel=1]{hyperref}

\newcommand{\CTE}{\alpha}
\newcommand{\CTEbar}{\bar{\CTE}}
\newcommand{\bk}{\beta}
\newcommand{\material}{Al--40Si}

\newcommand{\cooltemp}{$-196\,{}^\circ$C}

\newcommand{\sect}[1]{{section \ref{#1}}}

\newcommand{\subsect}[1]{{subsection \ref{#1}}}

\newcommand{\fig}[1]{{fig. \ref{#1}}}

\newcommand{\tab}[1]{{tab. \ref{#1}}}

\def\vct#1{\mathbf{#1}}

\allowdisplaybreaks

\begin{document}

\title{Theoretical compensation of static deformations of freeform multi mirror substrates} 

\newcommand{\affone}{{Fraunhofer-Institut f\"ur Angewandte Optik und Feinmechanik IOF, Leistungszentrum Photonik, Albert-Einstein-Stra{\ss}e\ 7, 07745 Jena, Germany}}
\newcommand{\afftwo}{{Friedrich-Schiller-Universit\"at Jena, Institut f\"ur Angewandte Physik IAP, Max-Wien-Platz~1, 07743 Jena, Germany}}

\author{Johannes Hartung}
\email{johannes.hartung@iof.fraunhofer.de}

\affiliation{\affone}

\author{Henrik von Lukowicz}
\affiliation{\affone}
\affiliation{\afftwo}

\author{Jan Kinast}
\affiliation{\affone}

\hypersetup{pdftitle={Deformation compensation},pdfauthor={Johannes Hartung et al.}}
\date{\today}

\begin{abstract}
Varying temperatures influence the figure errors of freeform metal mirrors by thermal expansion. 
Furthermore, different materials lead to thermo-elastic bending effects. 
The article presents a derivation of a compensation approach for general static loads. 
Utilizing perturbation theory this approach works for shape compensation of substrates which operate in various temperature environments.
Verification is made using a finite element analysis which is further used to produce
manufacturable CAD models. The remaining low spatial frequency errors are deterministically 
correctable using diamond turning or polishing techniques.\\
\vspace{0.1cm}
(120.4880) Optomechanics, (120.6810) Thermal effects, (260.3910) Metal optics, (120.4610) Optical fabrication, (220.1920) Diamond machining
\end{abstract}

\maketitle

\section{Introduction and Motivation}
Metal optics for space and ground-based applications have been state of the art for several years 
\cite{Riedl:2001, 
Gardner:Mather:Clampin:others:2006, 
Steinkopf:Gebhardt:Scheiding:others:2008, 
Risse:Gebhardt:Damm:others:2008,
Horst:Tromp:Haan:others:2008,
Nijkerk:vanVenrooy:vanDoorn:others:2012, 
Kerridge:Hegglin:McConnell:others:2012, 
Peschel:Damm:Scheiding:others:2014, 
Beier:Hartung:Peschel:others:2014,
Schuermann:Gaebler:Schlegel:others:2016}.
Polishable coatings like X-ray amorphous electroless nickel-phosphorus (NiP) %
\cite{Qin:2011, 
Park:Lee:1988, 
Hur:Jeong:Lee:1990, 
Dini:1991,
Syn:Dini:Taylor:others:1985,
Hibbard:1997,
Steinkopf:Gebhardt:Scheiding:others:2008,
Pramanik:Neo:Rahman:others:2008,
Carrigan:2011,
Feigl:Perske:Pauer:others:2015,
Gebhardt:Scheiding:Kinast:others:2011,
Risse:Gebhardt:Damm:others:2008, 
Gebhardt:Kinast:Rohloff:others:2014}
extend the range of applicable surface finishing methods to include chemical mechanical or computer controlled polishing (CMP, CCP) or 
Magnetorheological finishing (MRF) \cite{Steinkopf:Gebhardt:Scheiding:others:2008,
Schuermann:Gaebler:Schlegel:others:2016,
Carrigan:2011,
Feigl:Perske:Pauer:others:2015,
Gebhardt:Scheiding:Kinast:others:2011,
Risse:Gebhardt:Damm:others:2008, 
Gebhardt:Kinast:Rohloff:others:2014,
Kinast:Beier:Gebhardt:others:2015}
and improve the achievable optical surface accuracy.
There are several effects which lead to a change of the substrate shape (dimensional instability) and thus, of the optical surfaces. 
Dimensional instabilities are divided into three types: long-term, thermal cycling, and thermal \cite{Hibbard:1990, Folkman:Stevens:2002, Marschall:Maringer:1977, DeWitt:Eikenberry:Cardona:2008, Robichaud:Wang:Mastandrea:1998,
Kinast:Grabowski:Gebhardt:others:2014, Vukobratovich:Gerzoff:Cho:1997, Moon:Cho:Richard:2003, Wang:Shouwang:Jinlong:others:2016,
Zhou:Lin:Liu:others:2005}. 
Long-term instabilities are time dependent effects, like diffusion, phase transitions and stress relaxation of the materials used. 
Thermal cycling instabilities are plastic deformations due to thermal treatments. Different coefficients of thermal expansion (CTE) of the materials used 
lead to thermal instabilities. Dimensional instabilities lead in general to a non-acceptable degradation of the optical performance 
of the system which has to be restored by dimensional stable materials and one or more compensation approaches. Thermal instabilities 
can be compensated using kinematic means \cite{Yang:Liu:Wu:2012, Batteral:Fuss:Durieux:Martaut:2015}, active/adaptive optics 
\cite{Susini:Thomas:1993, Hubin:Noethe:1993, Reinlein:Appelfelder:Goy:others:2013, Goy:Reinlein:Devaney:others:2016, 
Reinlein:Brady:Damm:others:2016, Brady:Berlich:Leonhard:others:2017}, or compensation optics \cite{Krist:Burrows:1995, Slusher:Satter:Kaplan:others:1993}.
While all aforementioned compensation approaches take place in a post-manufacturing stage or during the operation of the
optical instrument, the subject of this article is compensation during manufacturing, possible when the static loads are known beforehand.

One important static effect is the isotropic linear thermal expansion or shrinkage of substrates made of one single material
under certain temperature loads.
For substrates coated with the previously mentioned polishing layer, non-linear and non-isotropic expansion and bending effects exist, too.
To minimize these effects, the mismatch of the CTE between the typical substrate material 
aluminum (e.g. Al6061) and the NiP coating -- which is responsible for the bending -- has to be reduced. Therefore
it is important to use a thermally matched material for the metallic substrate. In the case of NiP, 
40~w\% (weight percent) silicon-particle reinforced 
aluminum (designated as {\material} throughout the article) is considered, because the CTE mismatch between these materials is
$<0.5\times 10^{-6}\,\text{K}^{-1}$ \cite{Hibbard:1995, Hibbard:1990, Kinast:Hilpert:Lange:others:2014, Kinast:Hilpert:Rohloff:others:2014,
Rohloff:Gebhardt:Schoenherr:others:2010}.

Those effects lead to a specific surface form deviation (low spatial frequency contributions corresponding to figure errors), 
which reduces the performance of an optical system. 
In the case of a spherical mirror in a rotationally symmetric system, this performance degradation is predictable and correctable by adjusting the
distances along the optical axis in this system, but for a freeform mirror the thermal expansion 
leads to an irregular surface form deviation, which cannot be compensated easily anymore. Additionally, 
if there is a polishing layer applied to this metal mirror substrate, the bimetallic 
bending effect induces a freeform deviation of the mirror surface, even
for a relatively simple base shape. Even if the materials are thermally matched (small difference of their
corresponding CTEs at the respective operational temperatures), the occuring small mismatch produces an undesirable 
bending effect at large differences between manufacturing temperature and the well-defined operation temperature. 
To predict and correct these thermo-elastic surface form deviations (thermal instability) of the optical surfaces
before operation of the metal optics is the motivation of this paper. 

Within this article, the calculations to quantify these effects and to perform a compensation are carried out
for a metal two-mirror substrate which is part of a three-mirror-anastigmat (TMA). The two-mirror substrate consists of the primary mirror (M1)
and the tertiary mirror (M3). It is made of {\material} and coated with NiP.
Both mirrors M1 and M3 are aspheric with Zernike freeform contributions. For their manufacturing freeform techniques are necessary. 
The manufacturing and metrology usually
take place at room temperature, but the operational environment is different. E.g., infrared (IR) optics are used at {\cooltemp} to mitigate
thermal noise effects. This specific temperature is chosen because it is achievable by using liquid nitrogen and denotes its boiling temperature.
Although the mirror module is designed for the visual (VIS) spectral range, {\cooltemp} in comparison to room temperature defines the temperature 
load case throughout the present article \cite{DeWitt:Eikenberry:Cardona:2008, Fuentes:Manescau:SanchezDeLaRosa:1994,
Robinson:Huppi:Folkman:2002, Shen:Lin:Ma:others:1996}.

To describe the theory of thermo-elastic deformations, tensor calculus is necessary. 
All vectorial and tensorial calculations are carried out by utilizing a simplified Einsteinian summation convention, i.e. over double indices 
in one term is summed and the $\sum$ symbol is neglected (e.g. the matrix
multiplication $C_{ij} = A_{ik} B_{kj}$ is short for $C_{ij} = \sum_{k=1}^{3} A_{ik} B_{kj}$).
Further vector and tensor indices $i,j,k,\dots$ denoting their components are always from the middle of the Latin alphabet and their values are
$1,\dots,3$.
Different material indices $a, b, \dots$ are from the beginning of the Latin alphabet.
In the following calculations, the vector field $\vct{u}(\vct{x})$ denotes the deformation of a single volume element in a macroscopic
substrate at the field evaluation point $\vct{x} = (x_i) = (x, y, z)^T$. The dependency of certain fields on the point $\vct{x}$ is mostly omitted
within the formulas. 
Further $\sigma_{ij}~=~\sigma_{ji}$ is the stress tensor field and
$\delta_{ij}$ the Kronecker delta, which is $1$ for $i=j$ and $0$ for $i\ne j$, respectively. $\partial_i$ denotes the partial 
derivative in the $i$-th direction.
From the partial derivatives and the deformation field, the symmetrical infinitesimal geometrical linearized strain 
tensor $u_{ij}~=~\tfrac{1}{2}(\partial_i u_j~+~\partial_j u_i)$ is composed. Since the strain tensor is derived from the 
length of an infinitesimal distance between two points in the substrate before and after deformation, its full form is given by
$u_{ik} = \frac{1}{2}(\partial_i u_k + \partial_k u_i + \partial_\ell u_i \partial_\ell u_k)$, where the last term is neglected throughout the article.

However, the optical design of the example system is decoupled from the present article. The optical analyses, e.g.,
footprint export, and coordinate systems analysis were performed by using the program ZEMAX \cite{ZEMAX:2017}.
From ZEMAX, the surface exports need to be converted into a volume model. Therefore, the computer-aided design (CAD) program
CREO Pro/Engineer \cite{ProE:2017} is used. 
The finite element analyses (FEA) are carried out by using ANSYS \cite{ANSYS:2017} and the data analysis 
steps were done utilizing the computer algebra system Mathematica \cite{Wolfram:1991, Wolfram:2003}.

There are three main tasks to be achieved within this article:
\begin{enumerate}
 \item Theoretically derive general compensation formula to reduce optical influences of static load cases.
 \item Simplify compensation formula for thermo-elastic load case.
 \item Use finite element analyses to verify, apply, and compare different approaches for the derived compensation 
      formula for the M1M3 metal mirror substrate including polishing coating for the aforementioned thermo-elastic load case.
\end{enumerate}
Therefore, the present article is organized as follows. In \sect{sec:theoretical} the theoretical foundations of thermal 
effects in structural mechanics are
described and the compensation formula is derived. To put the theoretical derivation into an application context,
the TMA mirror system with the M1M3 mirror substrate and the corresponding manufacturing process chain are introduced in 
\sect{sec:example}.
The compensation formula as well as the model and manufacturing chain considerations are used in \sect{sec:fea} 
to pre-deform a model and to perform an FEA to obtain the residual deformation for the thermal load case mentioned. 
There are also some consistency calculations presented. The article concludes with presentation of the results 
of the FEA in \sect{sec:results} and shows a short outlook on the next steps in \sect{sec:summary}.

\section{Theoretical Analysis}\label{sec:theoretical}

\subsection{Elasticity Theory with Thermal Effects}\label{subsec:elasticitytheory}

From a theoretical point of view the top-down derivation starts by introduction of the total differential
of the inner energy with a strain energy contribution for the deformation of a volume body, 
\begin{align}
 \text{d}\mathcal{E} &= T\text{d}S + \sigma_{ij} \text{d} u_{ij}\,.
\end{align}
From that, it is possible to use a Legendre transformation and turn the expression of the inner energy
into one for the free energy,
\begin{align}
 \text{d}F &= -S \text{d}T + \sigma_{ij} \text{d} u_{ij}\,.
\end{align}
This leads to a thermodynamic expression for the stress tensor in 
the isothermic case of constant temperature $T$ \cite{Landau:Lifshitz:6:1991},
\begin{align}
 \sigma_{ij} &= \left(\frac{\partial F}{\partial u_{ij}}\right)_T\,.\label{eq:Fderivatives}
\end{align}
To get an expression for the stress tensor, it is necessary to provide a concrete form of the free energy $F$
in dependence of the strain tensor $u_{ij}$. The most simple way to construct such an expression is to use a series expansion in $u_{ij}$,
where the only appearing terms may be scalars built up from $u_{ij}$. It is further necessary that for $u_{ij}~=~0$ at
constant temperature, the stress tensor vanishes, too. The expansion has to start at quadratic order in $u_{ij}$. 
If $F$ consists of quadratic parts in $u_{ij}$ only, the arising dependency
between $\sigma_{ij}$ and $u_{ij}$ is linear and is called Hooke's law. 

Nevertheless, thermal expansion effects change the argumentation from above, since there is also
a net deformation without exterior forces acting on the material.
Therefore, the free energy of an isotropic single material deformable substrate for a small temperature range $\text{d}T$ starts at linear order
in $u_{ij}$ and is given by 
\begin{align}
	F(T) &= F_0(T) - 3 K(T) \CTE(T)\text{d}T\;u_{\ell\ell}\nonumber\\& + \mu(T)\;\left(u_{ik} - \frac{1}{3} \delta_{ik} u_{\ell\ell}\right)^2 + \frac{K(T)}{2}\; u_{\ell\ell}^2\,,
\end{align}
where $\CTE(T)$ (CTE), $K(T)$ (compression modulus), $\mu(T)$ (shear modulus) are considered to be spatially constant 
but functions of the temperature. 
For the present article, the temperature ranges are not small and in addition, 
the CTE data is obtained from a push--rod dilatometer, which works by applying a constant 
force to a sample and monitors its length change during well-defined temperature variations. Therefore, in this measurement all temperature 
dependencies of the elastic moduli are cumulated in an effective temperature dependency of $\CTE(T)$ and hence, they are 
taken as being constant, $K(T) = \text{const}$ and $\mu(T) = \text{const}$. 
Considering a finite temperature range $T-T_0$ for the elastic substrate, the free energy contribution 
$\sim\alpha(T)\text{d}T$ has to be integrated over this range. Since the temperature difference $T-T_0$ is a constant 
throughout this article, it is useful to substitute the integral by an averaged CTE,
\begin{align}
	\CTEbar &= \frac{1}{T - T_0} \int_{T_0}^T \text{d}T \CTE(T)\,.
\end{align}
This leads to a modified free energy for a finite temperature difference $T-T_0$
\begin{align}
	F(T) &= \bar{F}_0 - 3 K \CTEbar (T - T_0)\;u_{\ell\ell} \nonumber\\
	& + \mu\;\left(u_{ik} - \frac{1}{3} \delta_{ik} u_{\ell\ell}\right)^2+ \frac{K}{2}\; u_{\ell\ell}^2\,.
\end{align}
From the free energy the stress tensor $\sigma_{ik}$ can be obtained by \eqref{eq:Fderivatives},
\begin{align}
	\sigma_{ik} &= -3 K \CTEbar(T - T_0) \delta_{ik} + K\; u_{\ell\ell} \delta_{ik} + 2\mu\; \left(u_{ik} - \frac{1}{3} \delta_{ik} u_{\ell\ell}\right)\,.
\end{align}
The first term is the constant term due to linear contributions of $u_{\ell\ell}$ to the free energy.
The second and third terms are the contraction and shear contributions to the stress tensor, respectively.
As described in \cite{Landau:Lifshitz:6:1991} there may be no internal stresses in the material for a simple thermal expansion 
and therefore the stress tensor has to vanish, 
\begin{align}
	\sigma_{ik} &= 0\,.
\end{align}
This relation is valid for all field evaluation points $\vct{x}$ within the substrate. 
(In the present article, $\CTE$ is the 
linear coefficient of thermal expansion, while in \cite{Landau:Lifshitz:6:1991}, the volume coefficient of thermal expansion is used.
Further, the connection between shear modulus and contraction modulus on the one hand and Young's modulus and Poisson number, which are typically
used as material parameters, is $E~=~\frac{9 K \mu}{3 K + \mu}$ and $\nu~=~\frac{3 K - 2 \mu}{6 K + 2 \mu}$, respectively. 
If there are exterior forces acting on the substrate or stresses are induced, the full stationary equations of motion 
$-\partial_i \sigma_{ij} = f_j$ are to be solved.)
As it is known from representation theory, the trace part and the symmetric trace free (STF) or shear part transform 
independently under the group of the orthogonal $3\times3$ transformations with determinant one ($SO(3)$). 
Therefore, vanishing of the stress tensor leads to two conditions. 
From cancellation of the $\mu$ proportional STF part, it follows $u_{ik} = \text{const}\cdot \delta_{ik}$; from 
cancellation of the trace part, one obtains
\begin{align}
	u_{\ell\ell} &= 3 \CTEbar (T - T_0)\,.\label{eq:solstressfree}
\end{align}
Therefore,
\begin{align}
	u_{ik} &= \frac{1}{2} (\partial_i u_k + \partial_k u_i) = \CTEbar (T - T_0) \delta_{ik}\,.
\end{align}
For $T - T_0 = \text{const}$, one solution to this equation is 
\begin{align}
	u_k &= \CTEbar (T - T_0) x_k + u_{0\,k} + A_{km} x_m\,,\label{eq:divusolutionindex}
\end{align}
where $A_{km} = -A_{mk}$. In particular, this is valid for $A_{km} = \tfrac{1}{2}\varepsilon_{kjm} a_{0\,j}$.
The last part contributes to the rotation of $\vct{u}(\vct{x})$ only, but does not affect the symmetrical derivative $u_{ik}$.

Usually, for the deformation field there is no requirement to be zero at all of the volume boundaries of the solid body under consideration. This
would render the mechanical system over--constrained. Without any boundary conditions applied, the solution of \eqref{eq:divusolutionindex}
-- in classical notation -- is thus given by
\begin{align}
	\vct{u}(\vct{x}) &= \CTEbar (T - T_0) \vct{x} + \vct{u}_0 + \frac{1}{2} \vct{a}_0\times\vct{x}\,,\label{eq:divusolution}
\end{align}
where $\vct{u}_0$ and $\vct{a}_0$ are integration constants.

\begin{figure}[h]
\begin{minipage}{0.45\textwidth} 
\includegraphics[width=\textwidth]{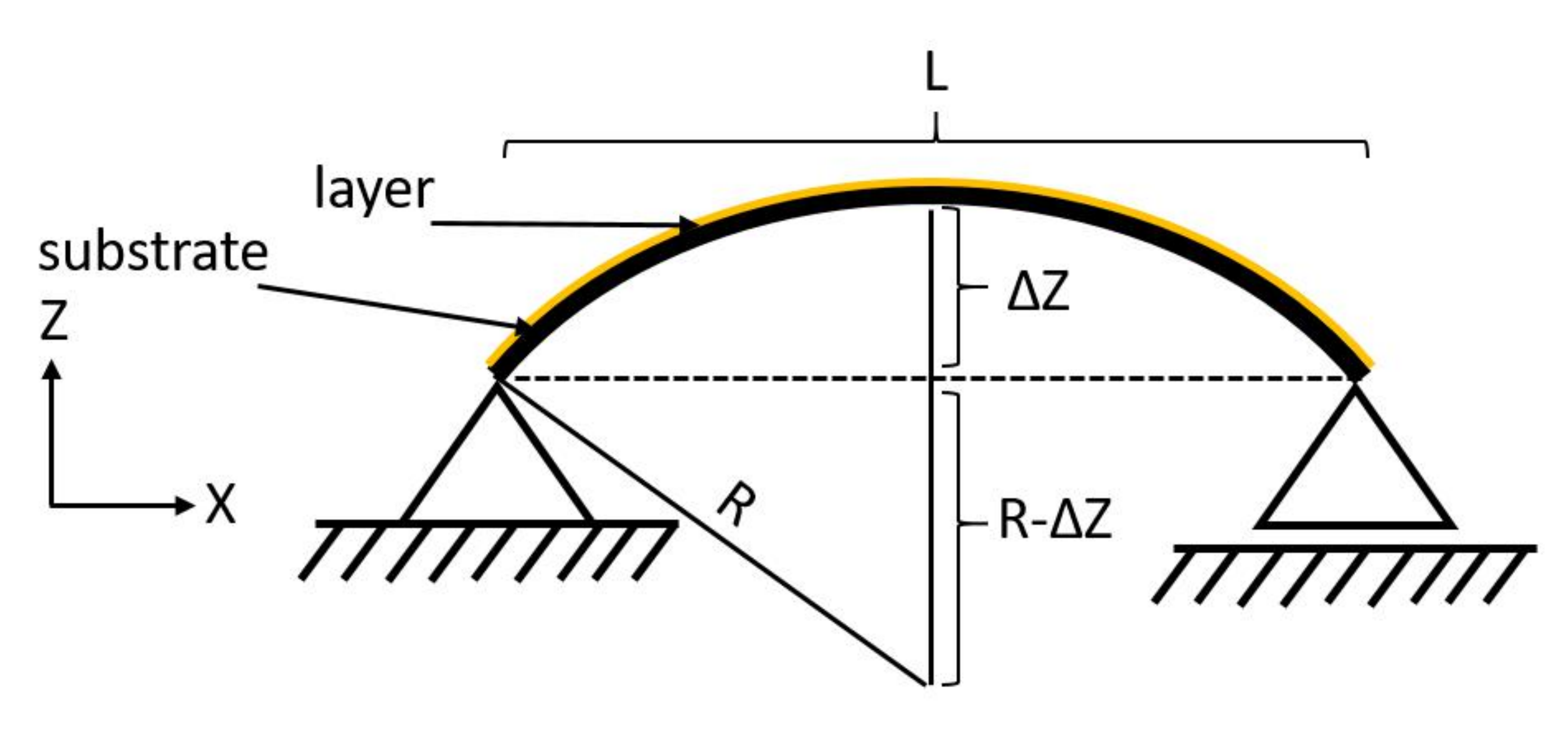}
\caption{\label{fig:bimetalbending}Principle of bimetallic bending effect}
\end{minipage}
\end{figure}

Until now, the thermal expansion of a substrate containing only one material is described. Equation \eqref{eq:divusolution} shows that
there are no bending effects as $\vct{u}$ depends only linearly on $\vct{x}$ only. If there are two materials present (e.g. NiP coated {\material}), 
the situation changes. Their different CTEs lead to different expansions and thus, stresses at the contact boundary and a bending of both occur. 
For two metal stripes bonded together with a respective thickness of $t_a$, Young's moduli of $E_a$, and averaged CTEs $\CTEbar_a$ ($a=1,2$), 
the analytical description of their curvature is given by
\begin{align}
	\frac{1}{R} &= 
		\frac{6(\CTEbar_1 - \CTEbar_2) (T - T_0)\left(1 + \tau\right)^2}
			{
				(t_1 + t_2) \left[3 
						\left(1 + \tau\right)^2 
						+ \left(1 + \tau \varepsilon\right)\left(\tau^2 + \frac{1}{\tau \varepsilon}\right)
				\right]
			}\,,\label{eq:bending}
\end{align}
where $\tau~=~t_1/t_2$ is the thickness ratio and $\varepsilon~=~E_1/E_2$ is Young's modulus ratio \cite{Timoshenko:1925}. 
There also exist several generalizations of \eqref{eq:bending} for, e.g., multi material stripes.
The deflection for a substrate of length $L$ fixed in $z$ direction on both ends is approximately derived from the geometrical relation
\begin{align}
\Delta z &= R\left(1 - \sqrt{1 - \frac{L^2}{4 R^2}}\right)\,,
\end{align}
which is shown in \fig{fig:bimetalbending}. The stripe is fixed in $z$ direction, but in $x$ direction it is only
fixed on the left hand side. Since the substrate length is mostly much smaller than the curvature radius $R$, one may
use a second order approximation in $L/(2 R) \ll 1$,
\begin{align}
\Delta z &\approx \frac{L^2}{8 R}\,.
\end{align}
In contrast for a one sided fixed stripe, which is free on the other side, the $z$ deformation is given by $\Delta z \approx \frac{L^2}{2 R}$.

Assuming that the corresponding deformation field $\vct{u}(\vct{x})$ is mostly dominated  by $u_z(x, 0, z)$ and the strip is fixed in $z$ direction at
both sides, $u_z(0, 0, z) = 0$, $u_z(L, 0, z) = 0$ and $u_z(L/2, 0, z) = L^2/(8 R)$, which leads in quadratic approximation to
\begin{align}
 \vct{u}_\text{quadratic}(x, 0, z) &\approx - \frac{1}{2 R} \left[\left(x - \frac{L}{2}\right)^2 - \frac{L^2}{4}\right] \vct{e}_z\,,\label{eq:ubendingstripe}
\end{align}
where $1/R$ from \eqref{eq:bending} is proportional to $T-T_0$ and the averaged CTE mismatch $\CTEbar_1 - \CTEbar_2$.

Although the formulas given in the last paragraphs are correct in the most simple cases only (i.e. no gradients for the thermal expansion
and metal stripe for the bending effects), they give an impression about the order of magnitude of certain effects ($\sim \CTEbar (T-T_0)$ for thermal 
expansion and $\sim (\CTEbar_1 - \CTEbar_2) (T-T_0)$ for the bending) and provide a reason for dividing the deformation field
$\vct{u}(\vct{x})$ the way it will be done in the following sections. Further, the M1M3 metal mirror substrate is neither a
simple block of material nor a metal stripe, but the formulas derived above lead to certain consistency checks verifying the FEA in 
\sect{sec:fea}.

\subsection{Deformation Compensation}\label{subsec:compensation}

\begin{figure}
\begin{minipage}{0.45\textwidth}
\includegraphics[width=\textwidth]{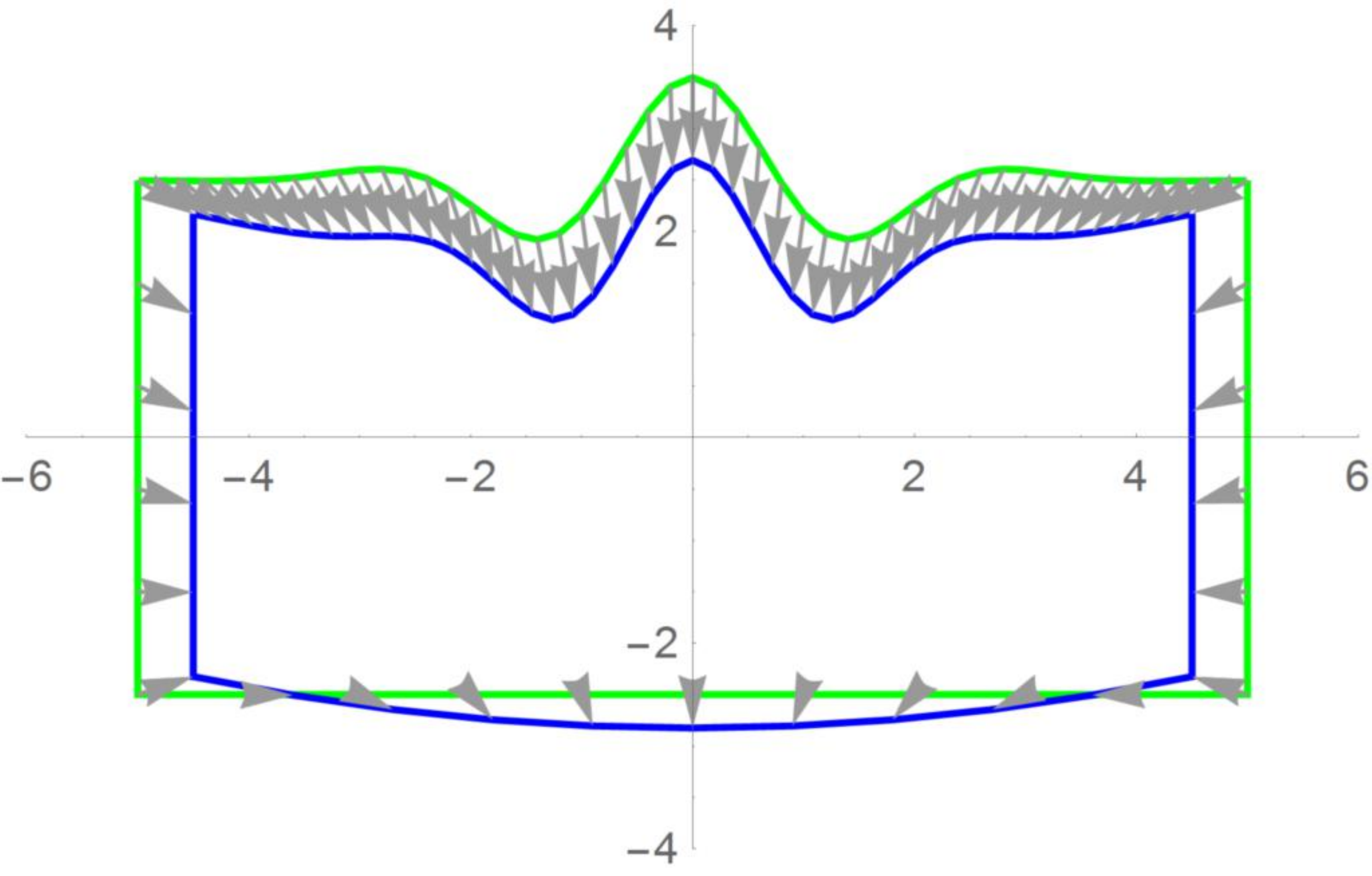}
\caption{\label{fig:udefall}
(a) Obtained cooling deformation. Green shows ideal substrate basis. Blue shows the deformed substrate after cooling.
}
\end{minipage}\hfill
\begin{minipage}{0.45\textwidth}
\includegraphics[width=\textwidth]{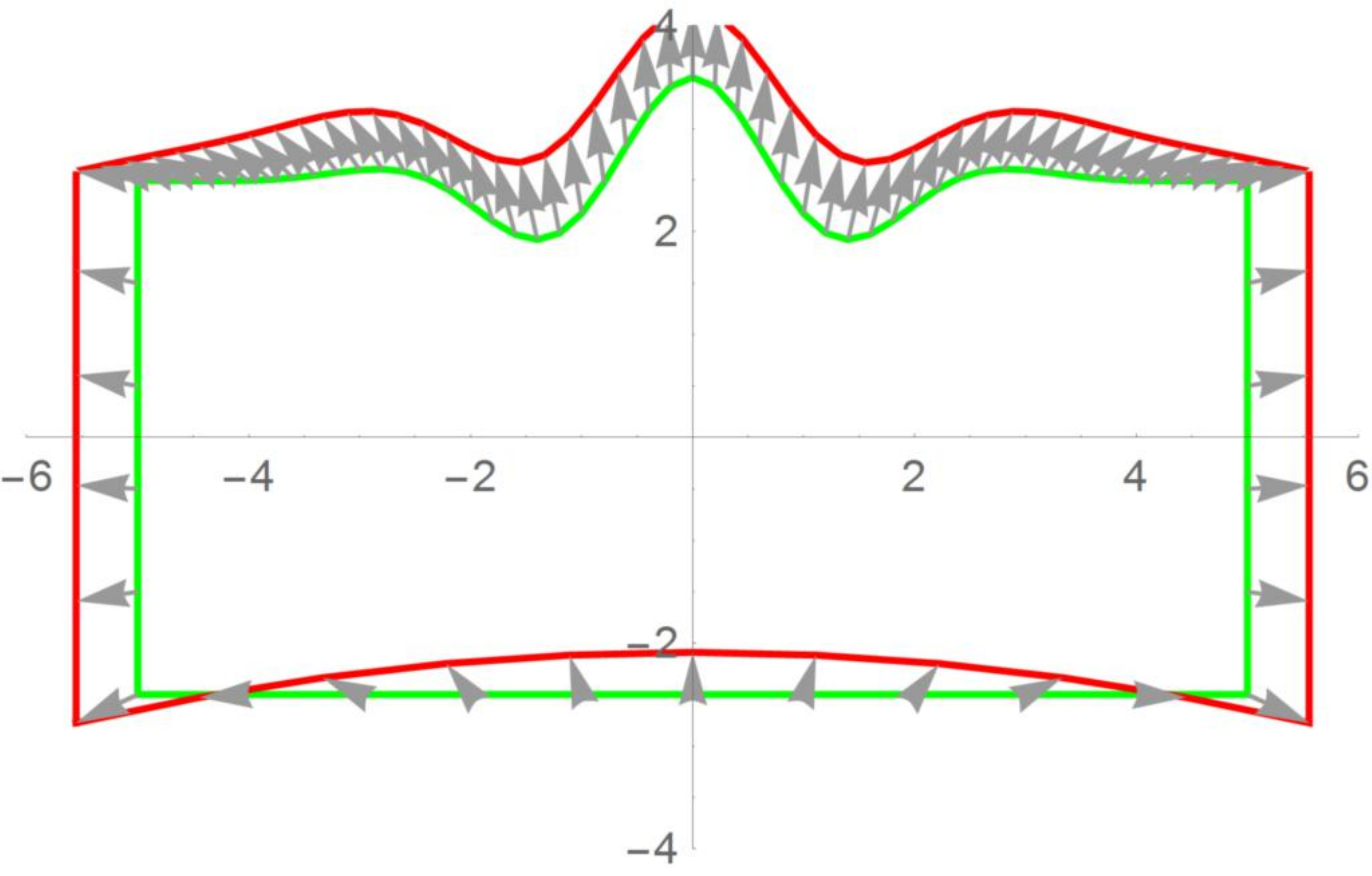}
\begin{flushleft} 
(b) Calculated cooling compensating ``heating'' deformation (red) and pre-deformed substrate such that its form corresponds
to the red line.
\end{flushleft}
\end{minipage}\hfill
\begin{minipage}{0.45\textwidth}
\includegraphics[width=\textwidth]{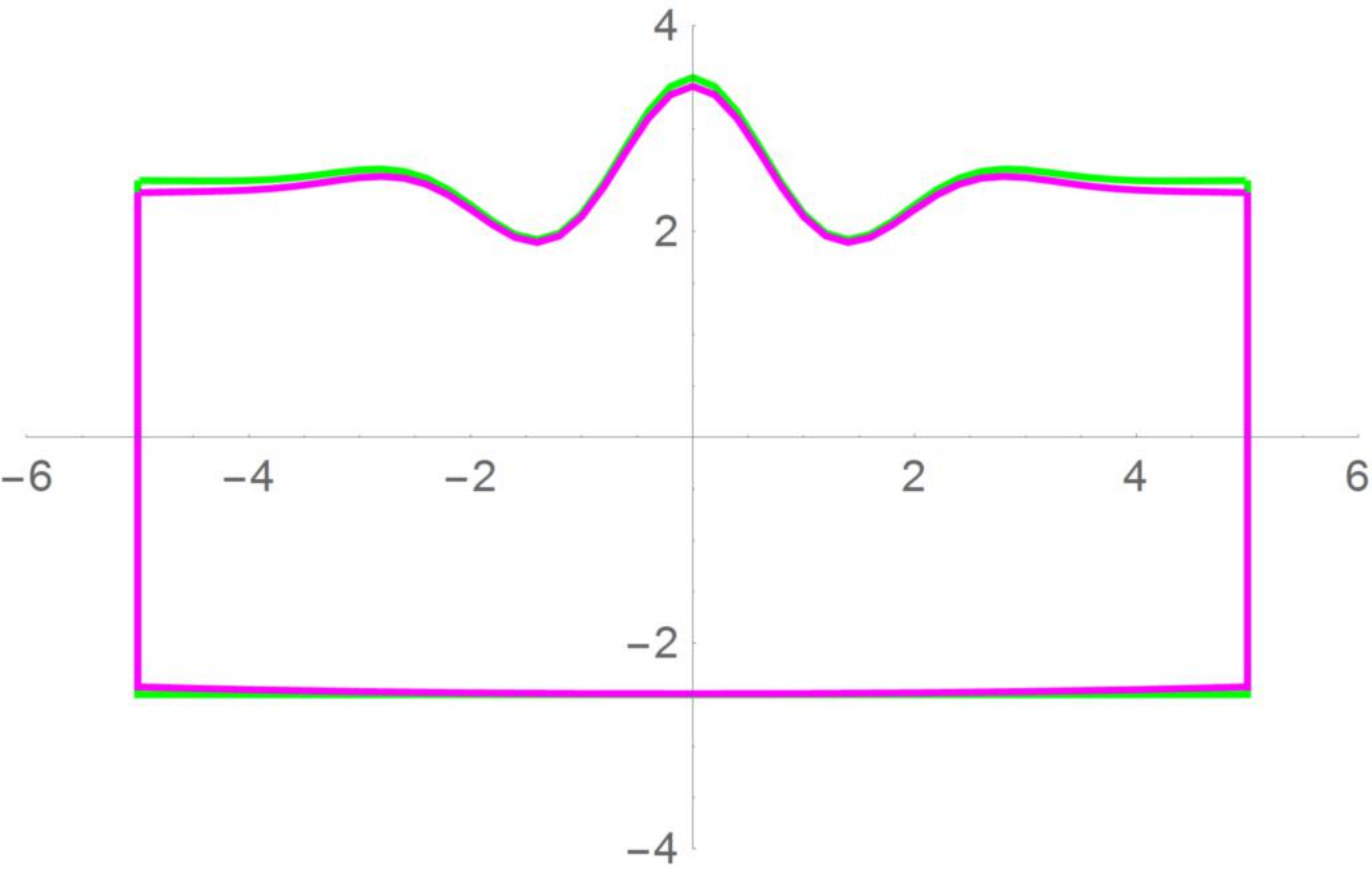}
\begin{flushleft}
(c) Cooling of pre-deformed (red) substrate should give a small residual deformation compared to the original (violet).
\end{flushleft}
\end{minipage}
\end{figure}

\newcommand{\uwo}[2]{{u_{(#1)#2\,w}}}
\newcommand{\uc}[1]{{u_{#1\,c}}}
\newcommand{\uw}[1]{{u_{#1\,w}}}

The results of \subsect{subsec:elasticitytheory} show that for the material combinations considered in the present article,
thermal expansion effects are much larger than thermal bending effects and therefore, $\vct{u}(\vct{x})$ 
is mostly affine linear with a small non-linear correction, namely
\begin{align}
	u_i &= D_{ij} x_j + b_{i} + u_{i\,\text{nl}}(x_k)\,.\label{eq:linuapprox}
\end{align}
The reason is, in \eqref{eq:divusolution} the field evaluation point $\vct{x}$ occurs in a linear manner only and
in, e.g., \eqref{eq:ubendingstripe} quadratic terms appear. The affine linear structure of the first two terms in \eqref{eq:linuapprox} is 
chosen, because a certain mounting of the M1M3 mirror module throughout the FEA may produce residual linear contributions in $\vct{u}(\vct{x})$, which cannot
be modelled by a simple divergence. To be able to extract those mounting effects, the full linear dependence is obtained by using a fit, whereas
the CTE is given by the divergence of the linear part of the vector field (divergence parts coming from the non-linear correction are neglected), 
\begin{align}
 \CTEbar (T - T_0) &= \frac{1}{3} D_{ii}\,,
\end{align}
and thus, by the trace of the matrix $D_{ij}$.
Besides, the rotation of the vector field can be expressed by the matrix components,
\begin{align}
 (\text{rot}\,\vct{u})_i &= \varepsilon_{ijk} \partial_j u_k = \varepsilon_{ijk} D_{jk} = a_{0\,i}\,.
\end{align}
The residual components are the symmetric traceless components of the matrix $D_{ij}$. These correspond to pure shearing of the material. 
By analysis of the different linear parts and subtracting them from the deformation field, the purely non-linear deformations are obtained. 
Therefore different parts of the deformation are well-defined and can easily be extracted from the fit data.

For a general derivation of the compensation procedure, a non-restricted $\vct{u}(\vct{x})$ is considered. The different steps of the compensation
procedure are shown in \fig{fig:udefall}. These steps are shown more explicitly throughout
the next paragraph. In this context, the article refers to cooling for the load case and heating for a virtual load case compensating
the deformation of the former one.

When a substrate is being cooled down during a finite element analysis from $T_0$ to operation temperature $T<T_0$ (see \fig{fig:udefall}(a)) 
its new nodal positions are given by
\begin{align}
	\vct{x}_c &= \vct{x} + \vct{u}_c(\vct{x})\,,
\end{align}
where $\vct{x}$ corresponds to the original nodal positions and $\vct{u}_c(\vct{x})$ is the deformation known according to the simulation. 
To compensate the known cooling deformation after applying the former load case it is necessary to deform the substrate before.
The deformation could be considered as a ``heating'' from operation temperature $T$ to $T_0$, see \fig{fig:udefall}(b). 
Its corresponding nodal deformation is denoted by
\begin{align}
	\vct{x}_w &= \vct{x} + \vct{u}_w(\vct{x})\,,
\end{align}
where $\vct{u}_w(\vct{x})$ is still unknown. The determination of $\vct{u}_w(\vct{x})$ for a known $\vct{u}_c(\vct{x})$ is the first goal of
the present article. Afterwards, the compensated substrate (pre-deformed by $\vct{u}_w$) is cooled down with the load case leading to $\vct{u}_c$ and
deformed nodal positions should end up near the ideal starting nodal positions $\vct{x}$, see \fig{fig:udefall}(c). 
In the following calculations, these considerations are formulated in a formal manner and a compensation equation is derived.
This means
\begin{align}
	\vct{x} &\stackrel{!}{=} \vct{x}_w + \vct{u}_c(\vct{x}_w) = \vct{x} + \vct{u}_w(\vct{x}) + \vct{u}_c(\vct{x} + \vct{u}_w(\vct{x}))\,.
\end{align}
Cancellation of $\vct{x}$ leads to
\begin{align}
\boxed{\vct{u}_w(\vct{x}) + \vct{u}_c(\vct{x} + \vct{u}_w(\vct{x})) \stackrel{!}{=} 0}\,,\label{eq:compensation} 
\end{align}
which is the compensation equation, where $\vct{u}_c$ is known and $\vct{u}_w$ should be calculated. 
Depending on the structure of $\vct{u}_c(\vct{x})$, this equation 
can be inverted analytically (e.g. for the purely linear case) or perturbatively (e.g. for a possible decomposition 
of $\vct{u}_c$ in some small and dominating part) or only numerically (for a general structure of $\vct{u}_c(\vct{x})$). 

Since the thermal load case is dominated by linear deformations,
\begin{align}
	\vct{u}_c &= D \vct{x} + \vct{b}\,,\label{eq:uclin}
\end{align}
\eqref{eq:compensation} can be solved exactly by
\begin{align}
	\vct{u}_{w\,\text{lin}} &= -(\mathds{1} + D)^{-1}(D \vct{x} + \vct{b})\,.\label{eq:uwlinapprox}
\end{align}

For performing a first order Taylor expansion in the non-linear correction, it is necessary to introduce a bookkeeping parameter $\bk$. 
The perturbative treatment is valid, because the non-linear correction is small compared to the linear part for thermal deformation 
problems. This leads to a correction of \eqref{eq:uclin} which is determined by a higher order polynomial fit
\begin{align}
	\vct{u}_c(\vct{x}) &= D\vct{x} + \vct{b} + \bk \vct{u}_{c\,\text{nl}}(\vct{x})\,.\label{eq:compensationbk}
\end{align}
Besides, the compensation field $\vct{u}_w = \vct{u}_{w\,\text{lin}} + \bk \vct{u}_{w\,\text{nl}}$ is divided into a linear and non-linear part.
After inserting these two splittings into \eqref{eq:compensation}, a lengthy expression arises:
\begin{align}
	&\vct{u}_{w\,\text{lin}}(\vct{x}) + D \vct{x} + D \vct{u}_{w\,\text{lin}}(\vct{x}) + \vct{b}\nonumber\\&+ \bk[\vct{u}_{w\,\text{nl}}(\vct{x}) + D \vct{u}_{w\,\text{nl}}(\vct{x})\nonumber\\&\quad +\vct{u}_{c\,\text{nl}}(\vct{x} + \vct{u}_{w\,\text{lin}}(\vct{x}) + \bk \vct{u}_{w\,\text{nl}}(\vct{x}))] \stackrel{!}{=} 0\,.
	\label{eq:compensationbkinserted}
\end{align}
In the Taylor expansion around $\bk=0$, all terms higher than linear order in $\bk$ have to be neglected to be consistent in the approximation.
By setting $\bk=0$, the purely linear part (see \eqref{eq:uwlinapprox} and \eqref{eq:uclin}) is restored. All other orders in $\bk$ are calculated by
sorting the Taylor expansion of \eqref{eq:compensationbkinserted} by powers of $\bk$ and setting them to zero term by term. Therefore, the leading order $\bk$ 
non-linear contribution of the compensation field is given by
\begin{align}
	\vct{u}_{w\,\text{nl}}(\vct{x}) &= -(\mathds{1} + D)^{-1} \vct{u}_{c\,\text{nl}}((\mathds{1} + D)^{-1}(\vct{x} - \vct{b}))\,,\label{eq:uwfirstnl}
\end{align}
and the whole compensation field including the linear part by
\begin{align}
	\vct{u}_{w}(\vct{x}) &\approx -(\mathds{1} + D)^{-1}\left[D \vct{x} + \vct{b} + \vct{u}_{c\,\text{nl}}((\mathds{1} + D)^{-1}(\vct{x} - \vct{b}))\right]\,.\label{eq:ulinnl}
\end{align}
The field $\vct{u}_w(\vct{x})$ shown in \eqref{eq:ulinnl} is referred as ``compensation'' deformation field.
It is a first order approximation to compensate a given $\vct{u}_{c}(\vct{x})$, if it can be decomposed into a dominant linear and a 
small non-linear part. In the following section, the example substrate M1M3 will be introduced and afterwards, the result \eqref{eq:ulinnl}
will be applied to the {\cooltemp} load case within finite element analysis.

\section{Example System and Process Chain}\label{sec:example}

\begin{figure}[h]
\begin{minipage}[t]{0.45\textwidth}
\includegraphics[width=\textwidth]{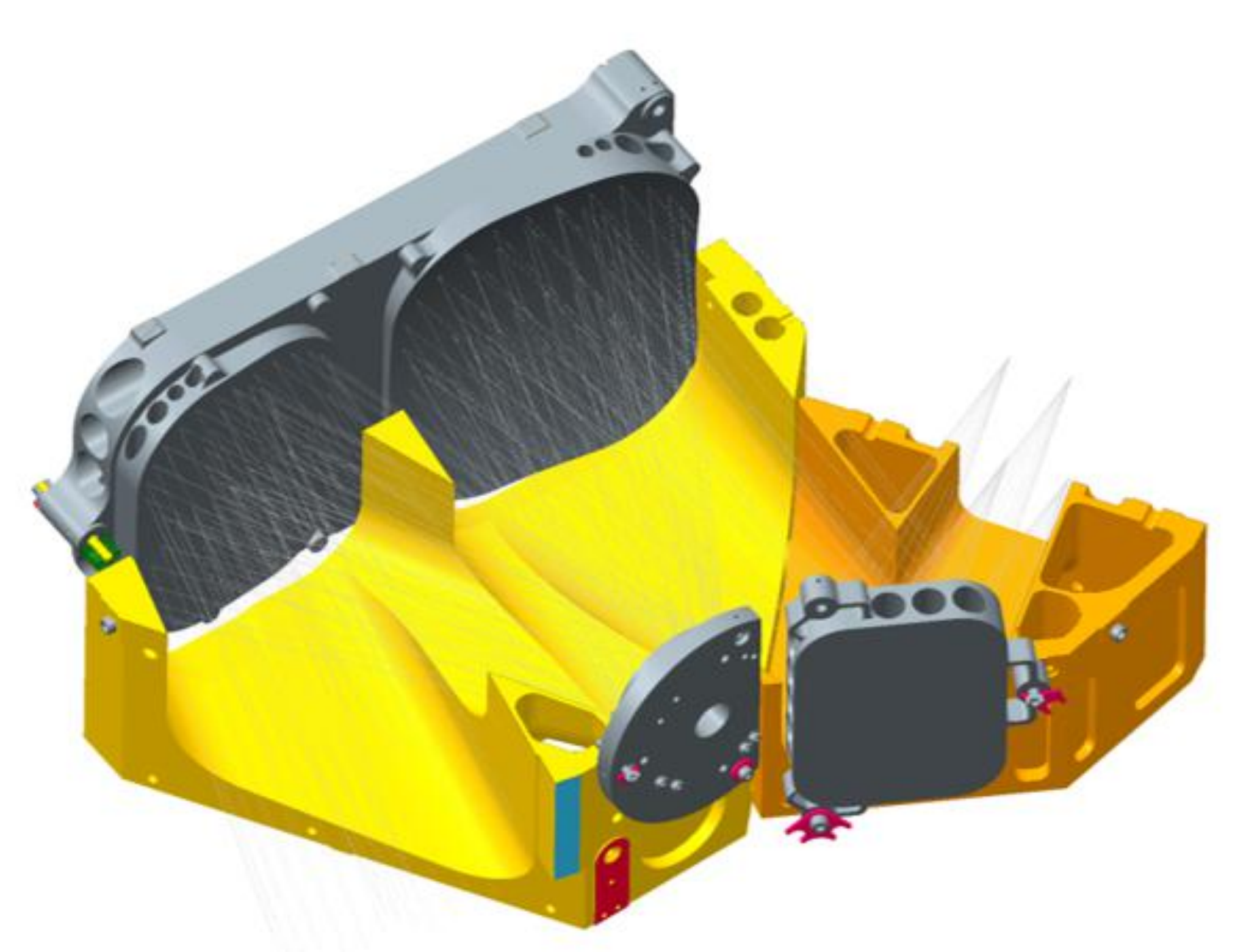}
\caption{\label{fig:SystemCADmodel}CAD model of the TMA with mirror substrates (grey) including housing (yellow).}
\end{minipage}\hfill
\begin{minipage}[t]{0.45\textwidth}
\includegraphics[width=\textwidth]{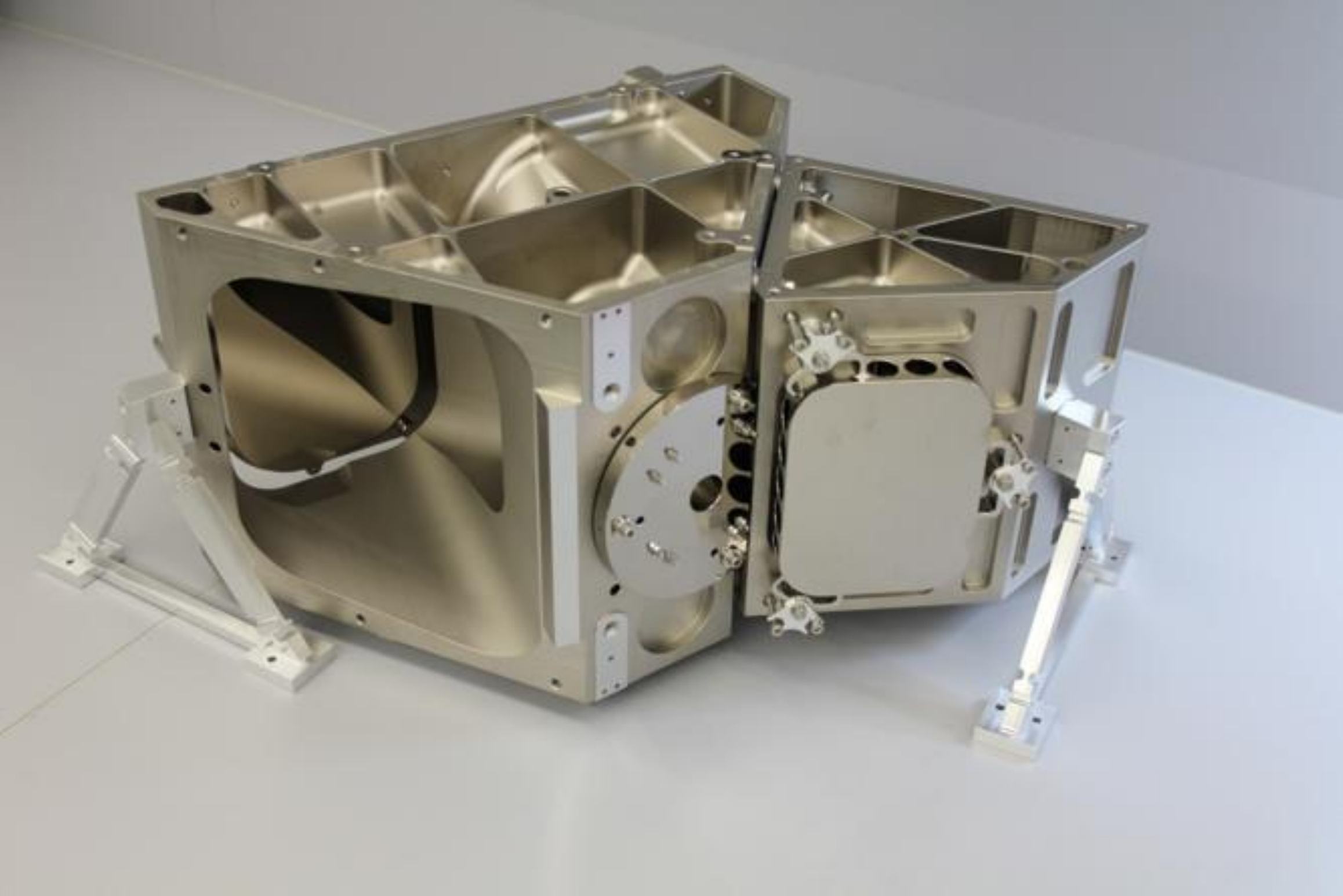}
\caption{\label{fig:SystemMounted}TMA after manufacturing and integration. The bipods shown in the picture are necessary to decouple the
system from the mounting plate.}
\end{minipage}
\end{figure}

\subsection{Optical Design Description of the TMA}\label{subsec:opticaldesign}

\begin{figure}[h]
\begin{center}
\includegraphics[width=0.45\textwidth]{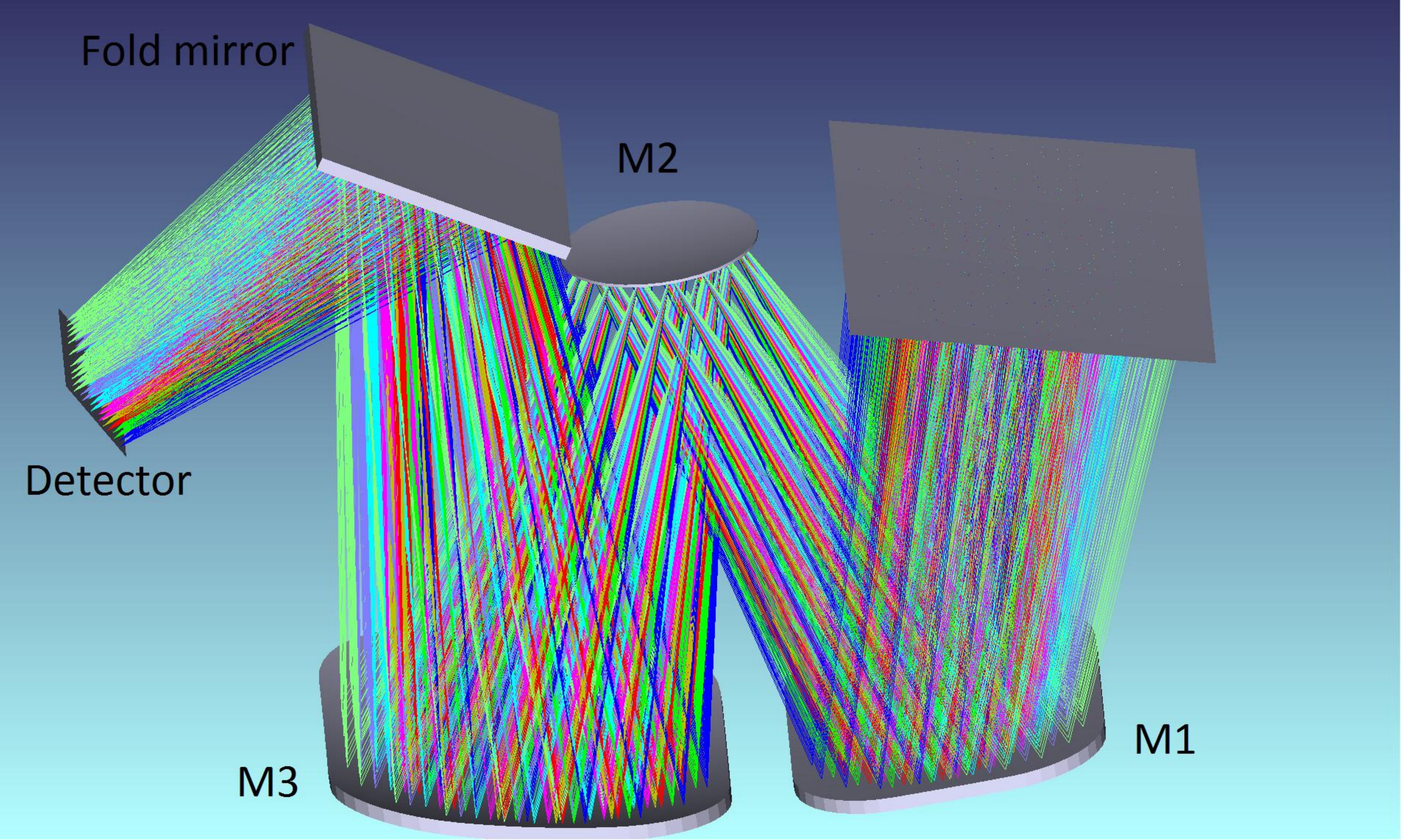} 
\caption{\label{fig:opticsdesign}3D optical design layout.}
\end{center}
\end{figure}

The optical system is a TMA (see \fig{fig:SystemCADmodel} and \fig{fig:SystemMounted}) 
consisting of three freeform mirrors and one folding mirror.
Except the folding mirror, the mirrors M1, M2, and M3 (see \tab{tab:opticsdata}) are referenced to a single axis and only
their $z$ position differs, see \tab{tab:opticaldesign} and \fig{fig:opticsdesign} respectively. 
Since M1 and M3 should be manufactured on one substrate
their respective optical coordinate system is also located at a coincident position with the
same axis orientations. The manufacturing coordinate system is then given by the fulfillment
of the manufacturing constraints. It is tilted and decentered in relation to the optical coordinate system. 
The surface M2 is located at the system stop. The entrance
pupil has a diameter of $81.25$ mm. Since the TMA is a telescope demonstrator,
the object plane is located at infinity. This means, each raybundle for different points of the field of view (FOV)
comes into the entrance pupil in different angles, but in a parallel manner. The FOV ranges between $-6.8^\circ$ and $6.8^\circ$ 
in $x$ direction and between $0^\circ$ and $6.3^\circ$ in $y$ direction.
Further, there is a focal plane array (FPA) located at the image plane. Since keystone and smile
distortion are typically specified for spectrometers they are also optimization criteria during 
the design process of this space technology demonstrator: its overall distortion magnitude is about $-2.94~\%$.

\begin{table}[h]
\center
\caption{\label{tab:opticaldesign}Optical design for the TMA mirror system. 
(Thickness $t$, decenter in $Y$ direction $DY$, tilt angle $RX$ around $X$ axis, clear aperture CA.)}
\begin{tabular}{lcccc}
\hline
	Surface & $t$ [mm] & $DY$ [mm] & $RX$ [${}^\circ$] & CA [mm${}^2$] \\
\hline
	Object	& $\infty$ & 0 & 0 & -- \\
	Entrance ap. & 23 & 0 & 0 & $109\times74.5$\\
	Coord. Break & 0 & 0 & $13.5$ & --\\
	M1 & $-220$ & 0 & 0 & user defined \\
	M2 & $220$ & 0& 0 & radius $40.5$ \\
	M3 & $-230$ & 0 & 0 & user defined \\
	Coord. Break & 0 & 0 & $-30$ & --\\
	Fold mirror & 0 & 0 & 0  &  $68.5\times54.5$\\
	Coord. Break & $220$ & 0 & $-30$ & --\\
	Coord. Break & $-30$ & $88.6$ & 0 & --\\
	Front surf. FPA & $30$ & 0 & 0 & radius $100$ \\
	Detector & 0 & 0 & 0 & $40\times20$ \\
\hline
\end{tabular}

\end{table}

\begin{table}[h]
\center
\caption{\label{tab:opticsdata}Optical design data for the mirrors of the TMA. The Zernike contributions were neglected.
Both vertex coordinate systems of M1 and M3 share a common point. The manufacturing coordinate system
(i.e. the center for the diamond turning, see \subsect{subsec:manufacturingchain}) 
is decentered and tilted vs. the optical coordinate system of the two vertices. The decenter in $y$ direction 
is $-15$ mm and the tilt around the $x$ axis is $-2.5^\circ$. (RoC $=$ radius of curvature, CC $=$ conic constant.)}

\begin{tabular}{lccc}
\hline
	Mirror & M1 & M2 & M3 \\
\hline
  RoC [mm] & $-1750.000$ & $-401.923$ & $-465.322$ \\
  CC [1] & $-69.850$ & $27.909$ & $-0.544$ \\
  A4 [mm${}^{-3}$] & $-3.422\times10^{-10}$ & $3.571\times10^{-7}$ & $-1.067\times10^{-9}$ \\
  A6 [mm${}^{-5}$] & $1.833\times10^{-14}$ & $-3.836\times10^{-10}$ & $-3.351\times10^{-15}$ \\
  A8 [mm${}^{-7}$] & $-7.748\times10^{-20}$ & $6.820\times10^{-16}$ & $-2.851\times10^{-20}$ \\
  Freeform & Zernike Noll & Zernike Noll & Zernike Noll \\
\hline
\end{tabular}
\end{table}

\subsection{Mechanical Design Description M1M3 Mirror Substrate}\label{subsec:mechanicaldesign}

The STEP surface export from ZEMAX is taken as a starting point for generating a volume CAD model of the different mirror
substrates. There are three different mirror substrates: the M1M3 substrate, the M2 substrate, and the fold mirror substrate.
Due to its complicated structure and to demonstrate the general usability of the algorithms developed, the M1M3 substrate
is considered as a demonstrator throughout this article. It does not have symmetries anymore, except the $YZ$ plane.
The mirror surfaces are aspheric with Zernike freeform contributions. Due to the deviation of the manufacturing coordinate system from the optical 
coordinate system, manufacturing requires freeform techniques, even for spherical symmetric surface shapes.
In this case, the compensation introduces also a freeform deviation from the optical design surface in general.

\begin{figure}[h]
\begin{minipage}[t]{0.45\textwidth}
\includegraphics[width=\textwidth]{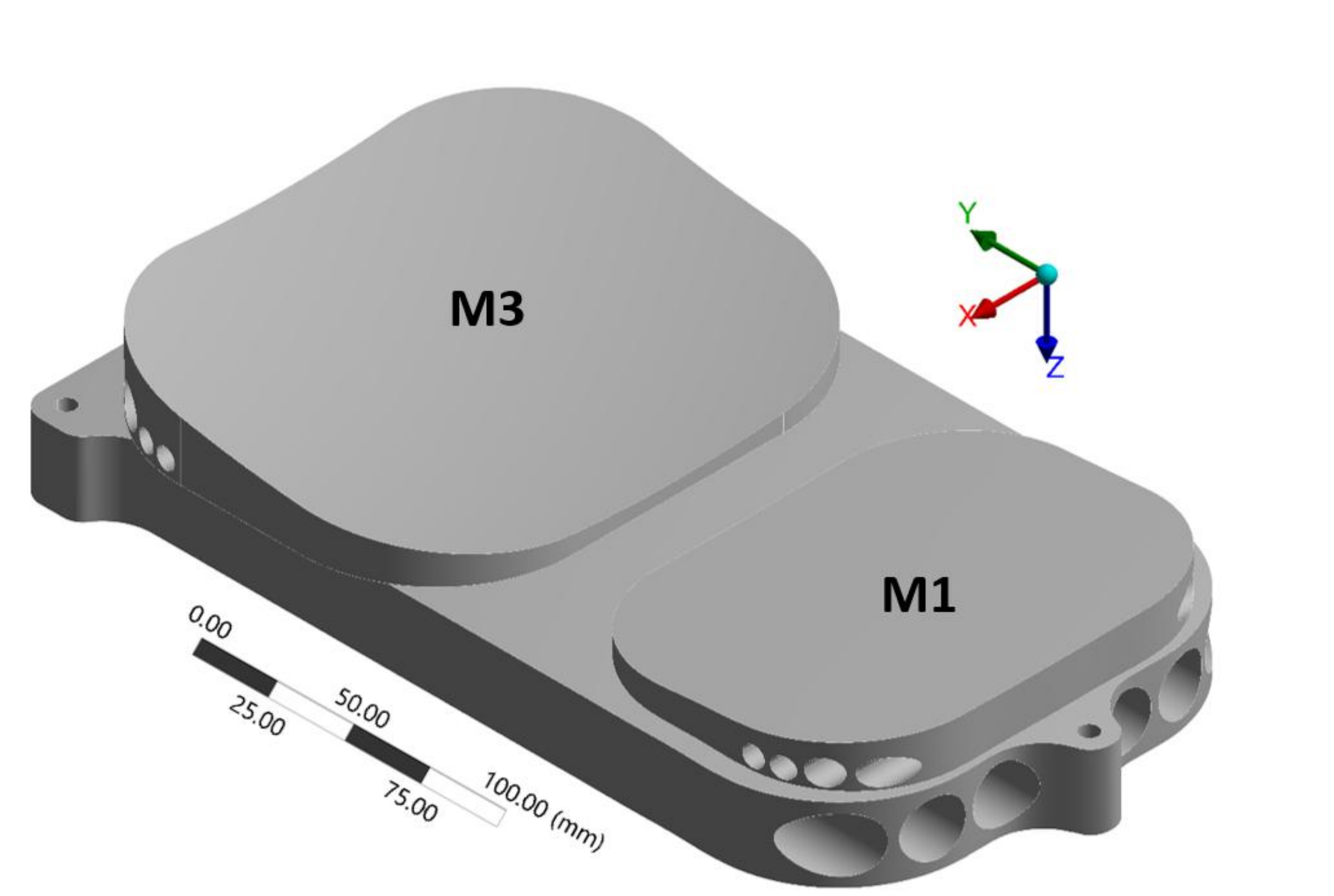}
\caption{\label{fig:CADmodel}CAD model of the mirror substrate. The bounding box of this model is 
$205\,\text{mm}\times339.5\,\text{mm}\times60\,\text{mm}$. The mass of the model manufactured from {\material} 
with $100\,\upmu$m NiP polishing coating is $4.06$ kg. $100\,\upmu$m NiP polishing coating thickness is chosen
to have enough material for the iterative UP diamond turning steps.}
\end{minipage}\hfill
\begin{minipage}[t]{0.45\textwidth}
\includegraphics[width=\textwidth]{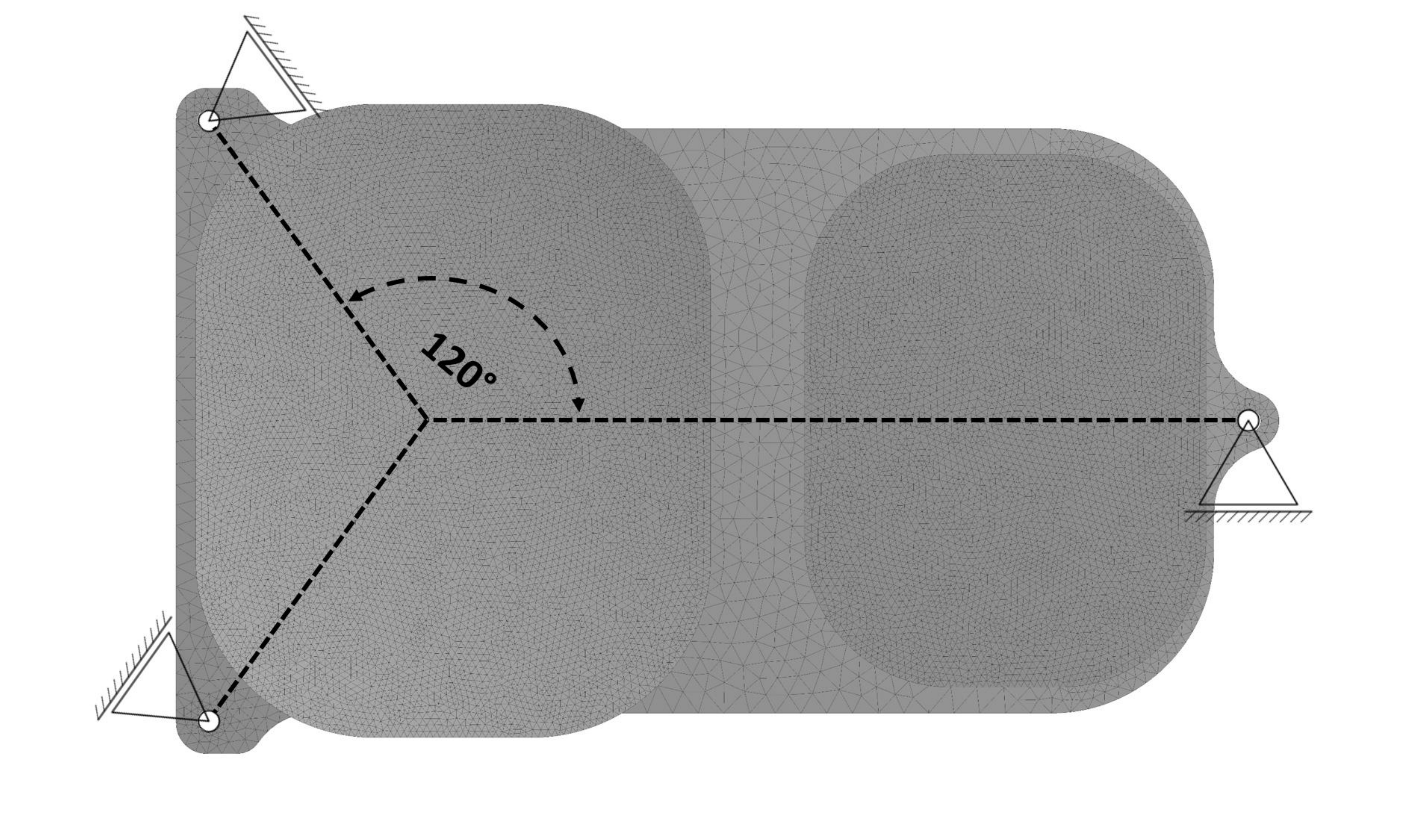}
\caption{\label{fig:isostaticmount}Isostatic mount points with an 120${}^\circ$ degree angle. The radial degree of freedom at the mountings 
is free, the tangential ones are fixed. The mesh elements are shown in color which represents the element quality, i. e. the aspect ratio.}
\end{minipage}
\end{figure}

\subsection{Overview Manufacturing Process Chain}\label{subsec:manufacturingchain}

The first step of manufacturing is an artificial aging of the rough part to minimize thermal cycling instabilities and to achieve a long-term stable
material behavior. Afterwards, the rough geometry of the substrate is obtained by a 
CNC manufacturing step, which is performed by a five-axis milling machine. Lightweight structures like crosswise bore holes are typically
drilled or electrical discharge machined. These steps depend on a ready-to-use CAD model, which needs to satisfy certain conditions to be manufacturable. 
Therefore, it is necessary to discuss possibilities to obtain a mostly simplified CAD model, 
see \subsect{subsubsec:cadcomp}.

\begin{figure}[h]
\begin{minipage}[t]{0.45\textwidth}
\includegraphics[width=\textwidth]{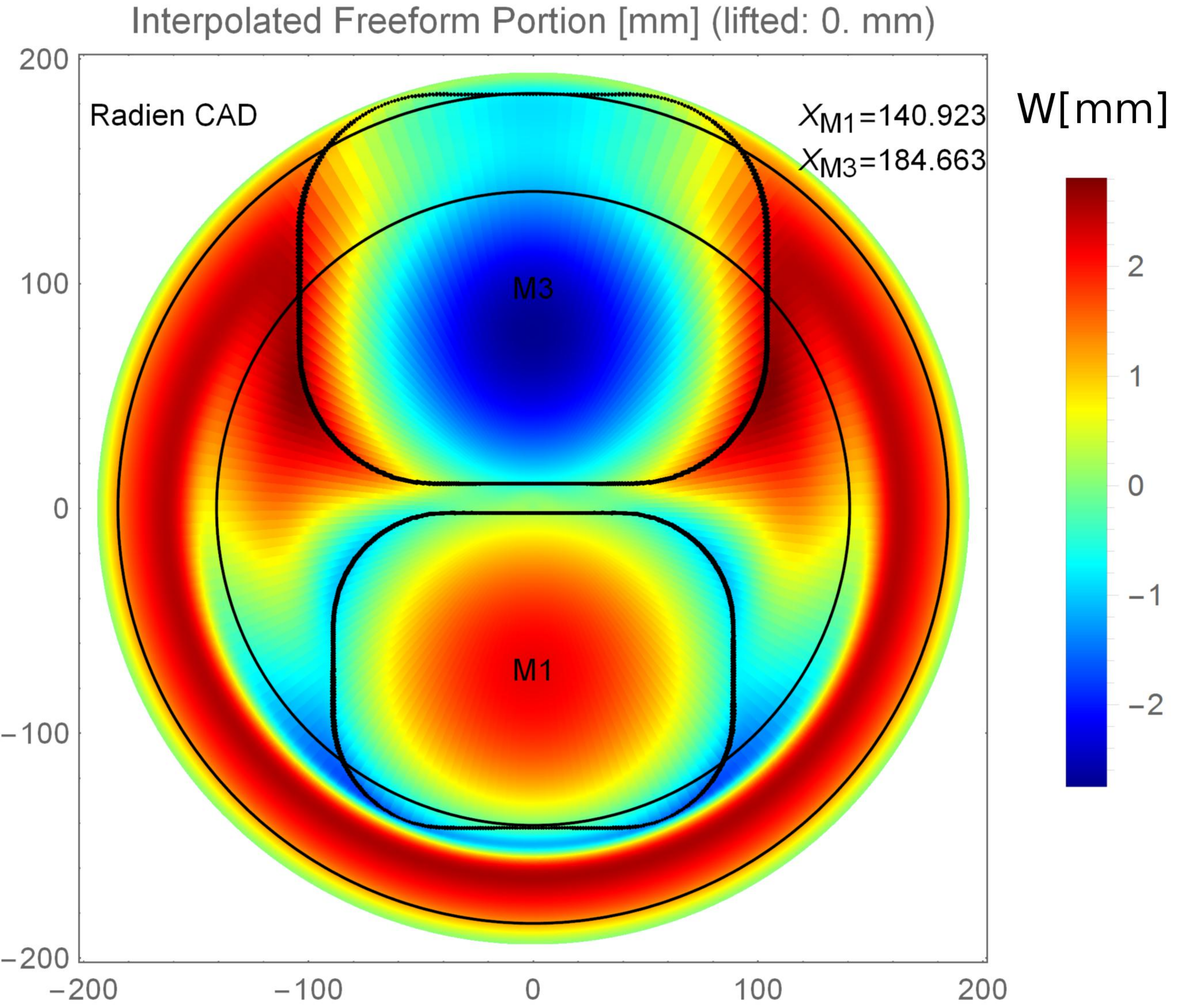}
\caption{\label{fig:toolpathsag}Freeform toolpath for fast tool servo with continuous toolpath between mirror surfaces M1 and M3.}
\end{minipage}\hfill
\begin{minipage}[t]{0.45\textwidth}
\includegraphics[width=\textwidth]{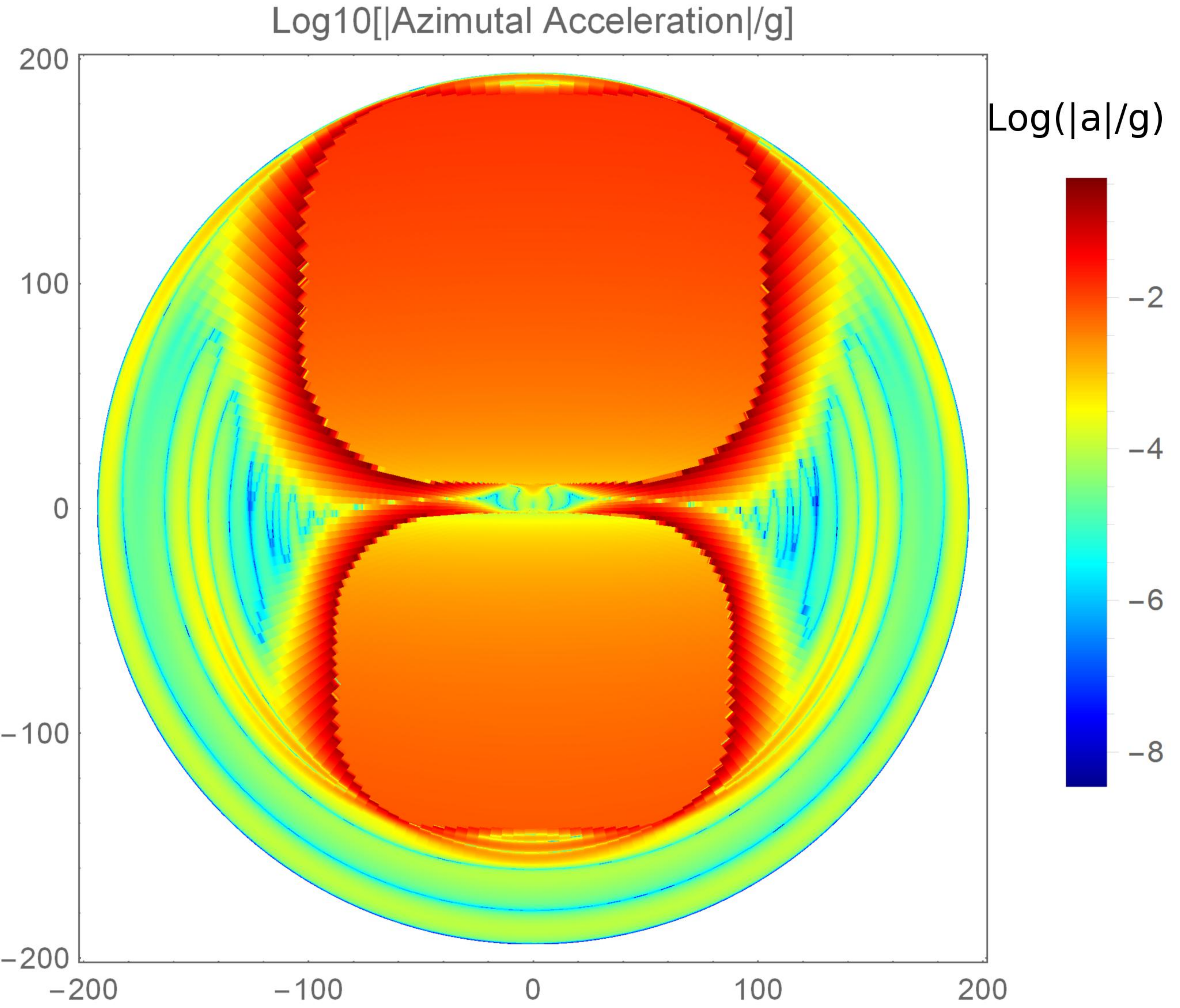}
\caption{\label{fig:toolpathacc}Tool acceleration for fast tool servo on a decadic logarithmic scale in units of $g = 9.81\,\text{m}/\text{s}^2$.}
\end{minipage}
\end{figure}

In the following, ultra-precise (UP) refers to an achievable figure error, which is in the micron range depending on the substrate size,
its deviation from rotational symmetry, its material, and several other parameters. For the UP manufacturing step, the following main processes are
available among others:
\begin{enumerate}
 \item Ultra-precise diamond turning for which a substrate is mounted on the turning machine ($C$-) axis  
	and a fixed tool is moved in a defined manner in $X$ and $Z$ direction to maintain the surface form. 
	For a freeform manufacturing using the slow tool servo, the $Z$ axis is controlled in dependency of 
	the $C$ values. For a higher dynamics of the freeform part,
	a fast tool servo is utilized, which introduces an additional voice coil driven $W$ axis. This axis has a
	lower maximal stroke than the whole $Z$ axis, but is more dynamical. It can only be used if the freeform part is not
	dominating the rotational symmetric part. For an imaging optical system, it is mostly possible to achieve this, see
	\fig{fig:toolpathsag} and \fig{fig:toolpathacc}, \cite{Beier:Hartung:Peschel:others:2014, Scheiding:Damm:Holota:others:2010, Risse:Scheiding:Gebhardt:others:2011}.
 \item Milling is necessary if the rotational symmetry of a part is broken strongly. It typically lasts longer than
      diamond turning processes. But $C$ axis milling in combination with diamond turning is also useful to provide references
      for tactile or interferometric metrology onto a metal mirror substrate. This is necessary to utilize the snap-together approach 
      \cite{Beier:Hartung:Peschel:others:2014}.
\end{enumerate}
A typical post-processing step to maintain a figure error down to $20$ nm r.m.s (root-mean-square). and around $1$ nm r.m.s. surface microroughness is the MRF step, where a fluid on a wheel and a magnetic field is used to generate a well-defined abrasive behaviour by controlling the dwell-time 
of the polishing wheel on the NiP coated substrate.

\section{Finite Element Analysis}\label{sec:fea}

\subsection{Model and Materials}

\begin{table}[h]
\centering
\caption{\label{tab:materialdata}Material data. The CTE mismatch is $0.5\times10^{-6} \text{K}^{-1}$.}
\begin{tabular}{lcc}
\hline
 Name & \material & NiP \\
\hline
 Young's modulus [MPa] & 102000 & 170000 \\
 Poisson number [1] & 0.27 & 0.27 \\
 CTE [K${}^{-1}\times 10^{-6}$] @ 20 ${}^\circ$C & 13.0 & 12.5 \\
\hline
\end{tabular}

\end{table}

The finite element model of the mirror substrate derived from the CAD model is built with 10-nodes tetrahedron (Solid187) 
elements for the substrate and 4-nodes quad (Shell181) elements for the NiP coating (see \tab{tab:materialdata} for the material data). 
Both elements have coincident nodes at the contact area. Shell elements for modelling the NiP coating are necessary, since 
for a sufficient mesh quality the aspect ratio of the linear dimensions of the volume elements should be of order one. This is 
violated for thin surface elements. (Since the thickness of the surface elements is approximately $100\,\upmu$m and their 
lateral dimensions are around $3\,$mm, they can be considered as thin.) Therefore, such coatings are usually modelled by 
using shell elements.

Based on the mounting concept 
for metallic mirrors, an isostatic mounting (ISM) is used for the boundary conditions model. 
Therefore, only the out-of-plane and the tangential component of the in-plane degrees
of freedom (DOF) are appended to three bearings, which have a symmetrical $120^{\circ}$ grid arrangement, 
see \fig{fig:isostaticmount}.
For this reason, a rigid body element (RBE3, not infinitely stiff) is attached at 
each of the three bores of the mirror substrate. This does not increase the stiffness of the model. 
The reference nodes (pilot node in ANSYS) of the RBE3 boundary condition 
are constrained in the tangential $\phi$ and $z$ DOFs with respect to a cylindrical coordinate system. In contrast, the
radial $r$ DOF of every RBE3 element is free in the mentioned cylindrical coordinate system. Therefore, the ISM generates 
a statically determined low-stress coupling with regard to thermal expansion. 

A thermal condition is applied for both the mirror substrate body and the
layer shells. Each of the elements composing the substrate and the
coating are set to a specific temperature $T \ne T_0$, justifying usage of
\eqref{eq:solstressfree} and \eqref{eq:linuapprox}
for modelling the $\vct{u}$ field without any NiP layer (e.g. Al6061 or {\material} only). 
For the model with NiP layer, there are stresses induced at the boundary of the substrate 
due to different CTEs and hence, non-linear corrections to \eqref{eq:linuapprox} 
have to be taken into account, see \eqref{eq:ulinnl}.

Modelling the $\vct{u}$ field in this context means that after FEA, a global fit of the deformation
field over the model is carried out by using \eqref{eq:linuapprox}. From there, the compensation field is determined and put into
another FEA with the same load cases. This leads to a net deformation on the mirror surfaces, which
are compared to the optical design afterwards. Notice that a fit in general violates the boundary conditions
of the FEA and therefore the compensation field does not satisfy any well-defined boundary conditions, too. 
However, this is acceptable, since it is only used for deformation of the base geometry.

\subsection{Simulations}\label{subsec:simulations}

\begin{figure}[h]
\begin{minipage}[t]{0.45\textwidth} 
\includegraphics[width=\textwidth]{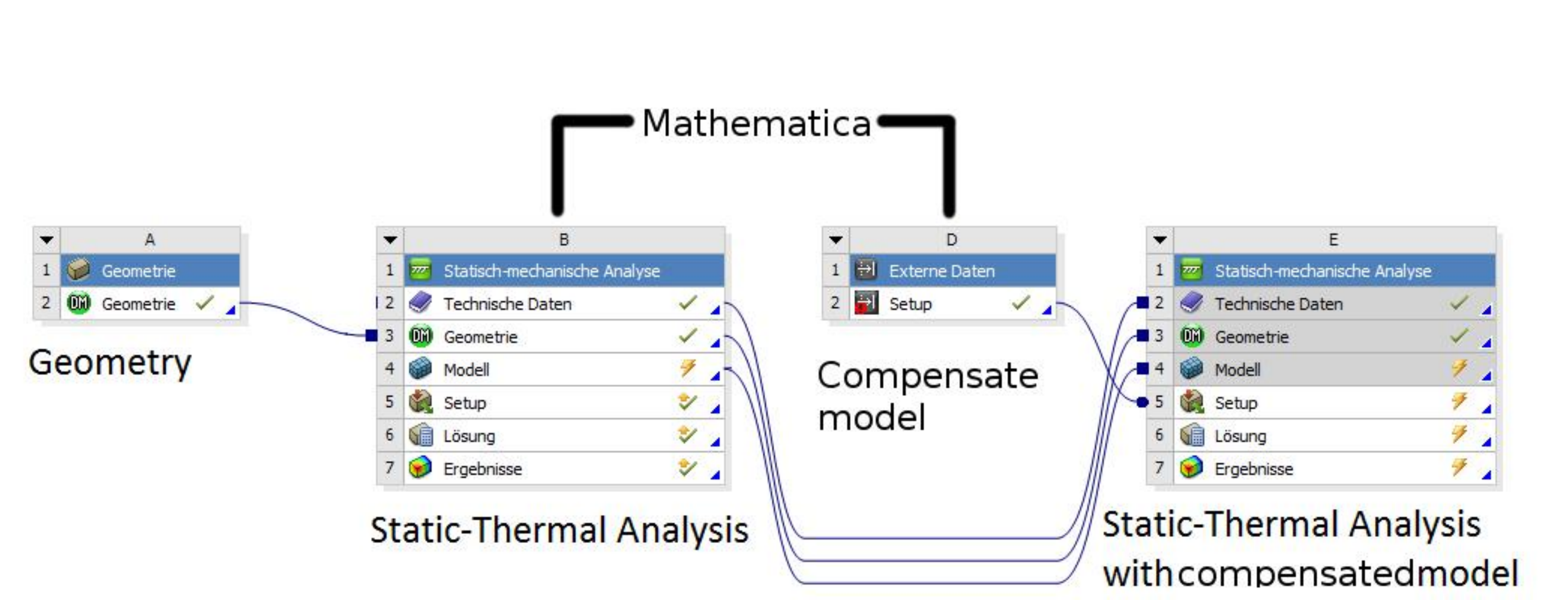}
\caption{\label{fig:flowdata}Data flow for exact compensation within the ANSYS FEA software. The load case is given by a temperature ramp starting at
$22\,{}^\circ$C and ending at \cooltemp.}
\end{minipage}\hfill
\begin{minipage}[t]{0.45\textwidth} 
\includegraphics[width=\textwidth]{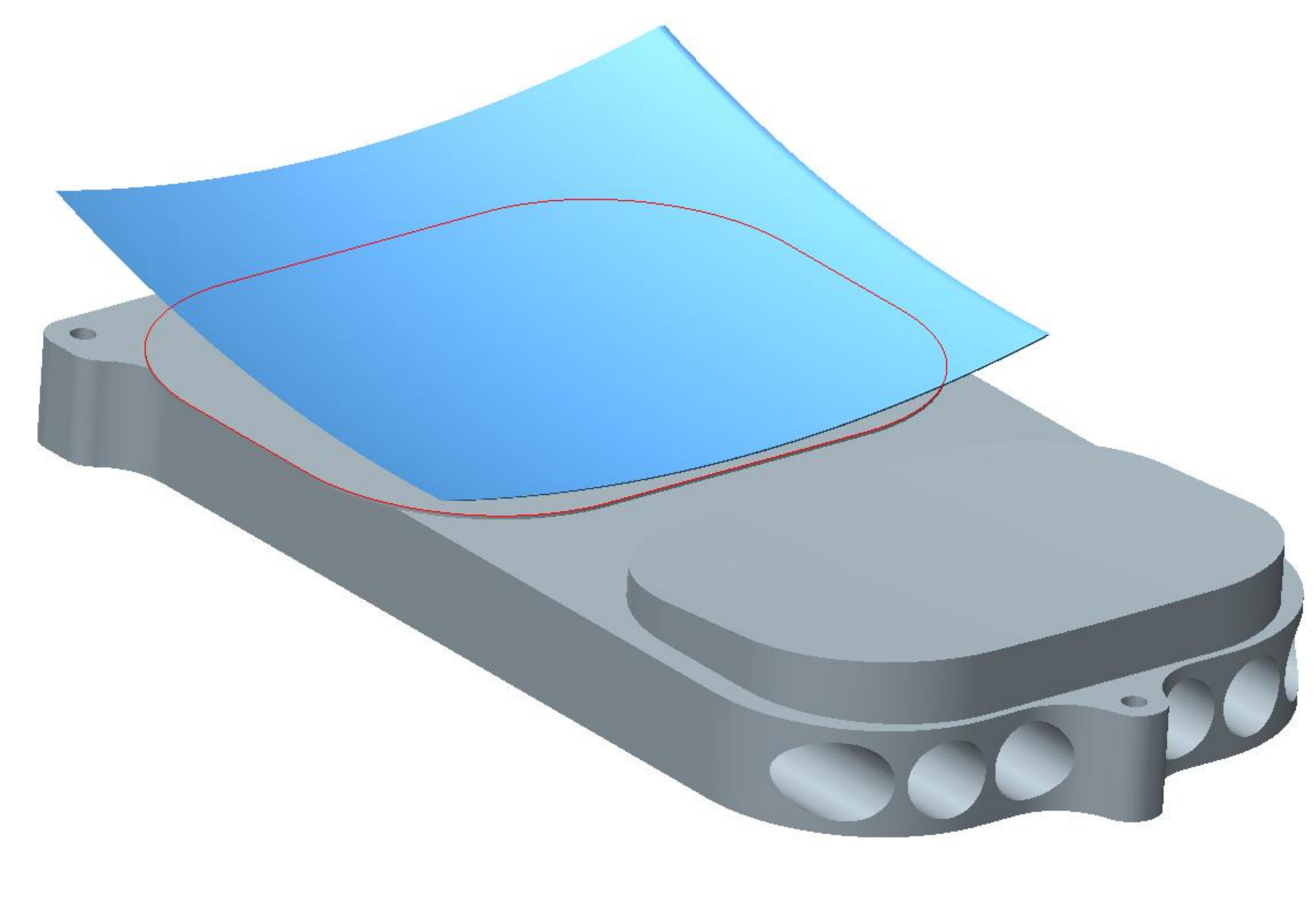}
\caption{\label{fig:CADmodel_surfaces}Compensation of the CAD model of the mirror substrate by generating the necessary exact compensated surfaces
and increasing the size of the substrate isotropically.}
\end{minipage}
\end{figure}

\subsubsection{Dataflow Exact Compensation}\label{subsubsec:exactcomp}

In the ``exact'' compensation, the field $\vct{u}_w$, see \eqref{eq:ulinnl}, is used to displace the nodal positions
in the whole model. As far as the deformation field has a smaller magnitude than the size of the elements, it does not lead to any problems
like breaking of elements. During mechanical design stage (see \subsect{subsec:mechanicaldesign}) a CAD model is created to
perform the CNC manufacturing, and also for the documentation of every step in the process chain. It is necessary to maintain the connection of 
the FEA model (mesh together with boundary conditions) to the CAD model after the compensation as well. By the calculated nodal deformation 
of the whole model, this connection is lost. Another major drawback is that edges may not be straight anymore or plane surfaces
may also be deformed. Therefore, the exact compensation of all nodal position is of theoretical interest, but does not satisfy certain
constraints, which are necessary for manufacturing.

\subsubsection{Dataflow CAD-based Compensation}\label{subsubsec:cadcomp}

Thus, a reduced compensation is performed. Starting from the fully deformed model, the necessary exact surface deviation 
is extracted. This is used to modify the former ideal surfaces. But for the volume model only an isotropic expansion is performed by using 
the effective CTE extracted from the $D_{ij}$ matrix fit. This leads to straight lines and surfaces after the deformation and therefore,
they are still usable for mounting and clamping purposes on the machines. In \fig{fig:CADmodel_surfaces}, building the CAD conform model
is shown. Exactly deformed surfaces are created by using spline patches, which were generated from the compensated nodal positions on the surface.
Those spline patches are used as a boundary surface for the extrusion of the surface footprints from the main substrate after its isotropic
expansion. 

In the following section, the resulting figure error after cooling the compensated geometry is compared for these two approaches.

\section{Results}\label{sec:results}

During temperature change, each metal undergoes a certain thermal expansion. This thermal expansion is isotropic and linear for a single 
isotropic material and is non-isotropic and non-linear for a group of bonded different materials. For the case of metal mirror substrates, 
both effects lead to a deviation of the optical surface shapes from the intended design form. Throughout the present article, those effects 
were discussed for a metal mirror substrate with two mirrors, which is part of a TMA introduced in \sect{sec:example}. The double mirror 
substrate and the thermal load case are both 
examples to demonstrate the feasiblity of the general compensation approach developed in \subsect{subsec:compensation} to compensate the 
deviations mentioned above. This approach relies on the precise knowledge of the temperature load, which arises between manufacturing process and 
operational environment of the optical component.

Given the compensation formula \eqref{eq:compensation}, there were two algorithms developed to change the substrate in such a manner that
a further application of the temperature load leads to a net surface contribution, which is far smaller than without compensation. Although
\eqref{eq:compensation} is exact and quite general, the decomposition of the thermal load in this article into a large expansion part 
and a small bending part allows to find a closed form first order perturbative compensation formula for the necessary deformation 
of the original substrate, see \eqref{eq:ulinnl}.

The exact compensation leads to a non-manufacturable geometry of the part, because, e.g., straight lined lightweight structures get bended in general.
This approach (see \subsect{subsubsec:exactcomp}) is not feasible for production. For the thermal load case, 
the deformations may be divided into a large linear and a small non-linear bending part. Therefore, it is also possible to produce a CAD model 
with the desired compensation properties, which is also manufacturable in the CNC fabrication step before ultra-precise diamond turning
(see \subsect{subsubsec:cadcomp}). Further, the $z$ value 
compensation of the optical surfaces is in a range, which is also tractable by diamond turning techniques. Therefore, the developed
compensation approaches are in-situ and fit into an existing manufacturing chain for metal mirrors right before the finishing process steps.

In the following, the optical surface deviations from design are illustrated and compared
between the uncompensated model, the exact nodal compensation approach and the CAD compensation one.

\newcommand{\resultcaption}[6]{{M{#1} after cooling, {#2}. Form deviation \mbox{p.-v.} $\approx~{#3}$~{#5}m, r.m.s. $\approx {#4}$ {#5}m. {#6}}}
\newcommand{\figwidth}{0.45}

\newcommand{\monepvnocomp}{21.8}
\newcommand{\monermsnocomp}{5.8}
\newcommand{\mthreepvnocomp}{34.4}
\newcommand{\mthreermsnocomp}{8.6}

\newcommand{\monepvCADcomp}{301}
\newcommand{\monermsCADcomp}{61}
\newcommand{\monefinalpvCADcomp}{166}
\newcommand{\monefinalrmsCADcomp}{19}
\newcommand{\mthreepvCADcomp}{618}
\newcommand{\mthreermsCADcomp}{160}
\newcommand{\mthreefinalpvCADcomp}{171}
\newcommand{\mthreefinalrmsCADcomp}{21}
\newcommand{\monepst}{$-2.37\,\upmu$m}
\newcommand{\monetlt}{$0.37''$}
\newcommand{\mthreepst}{$-2.36\,\upmu$m}
\newcommand{\mthreetlt}{$0.79''$}

\newcommand{\monepvexcomp}{118}
\newcommand{\monermsexcomp}{26}
\newcommand{\mthreepvexcomp}{97}
\newcommand{\mthreermsexcomp}{16}

\begin{figure}
\begin{minipage}[t]{\figwidth\textwidth}
\includegraphics[width=\textwidth]{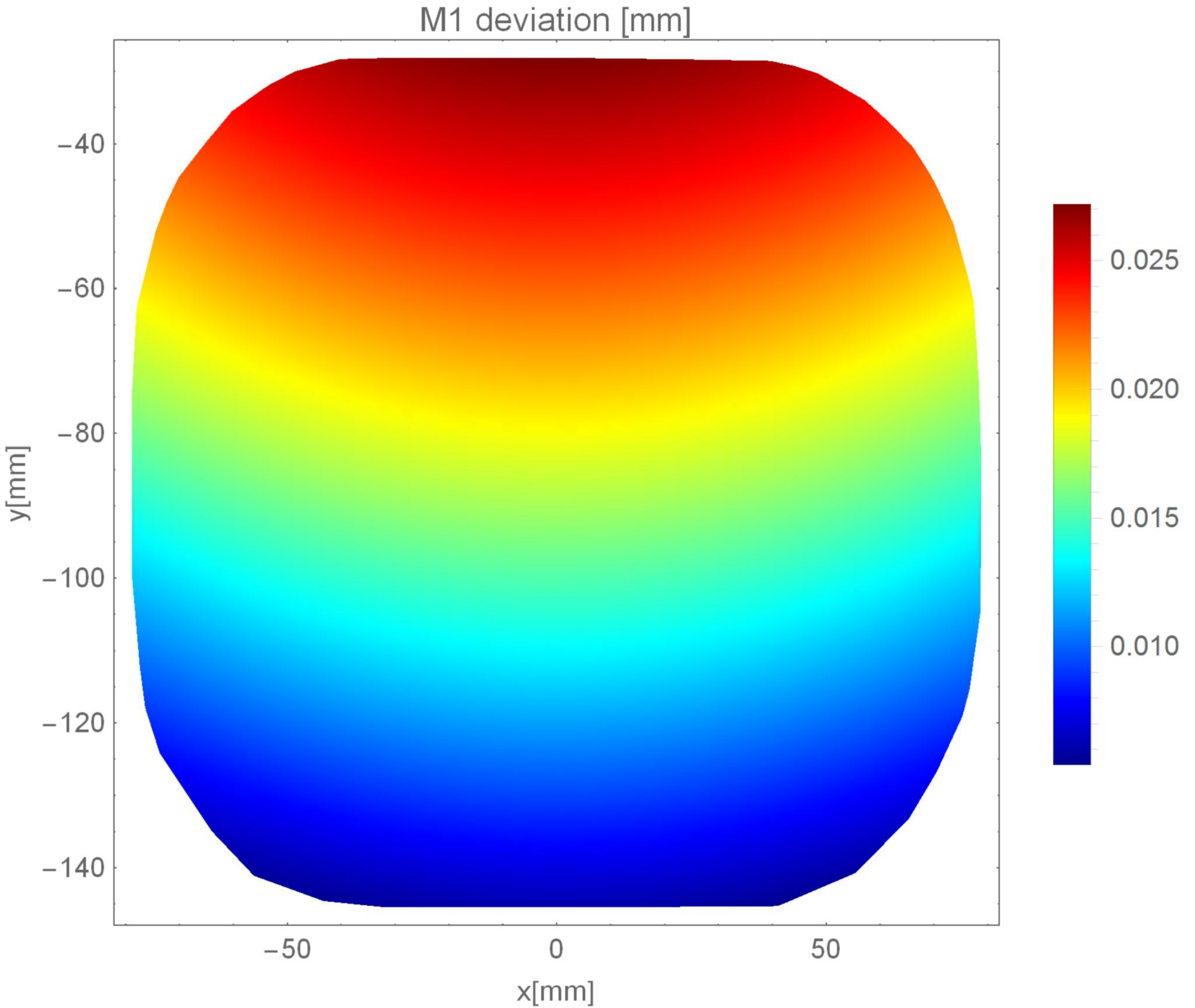}
\caption{\label{fig:m1nocomp}\resultcaption{1}{uncompensated}{\monepvnocomp}{\monermsnocomp}{$\mu$}{}} 
\end{minipage}\hfill
\begin{minipage}[t]{\figwidth\textwidth}
\includegraphics[width=\textwidth]{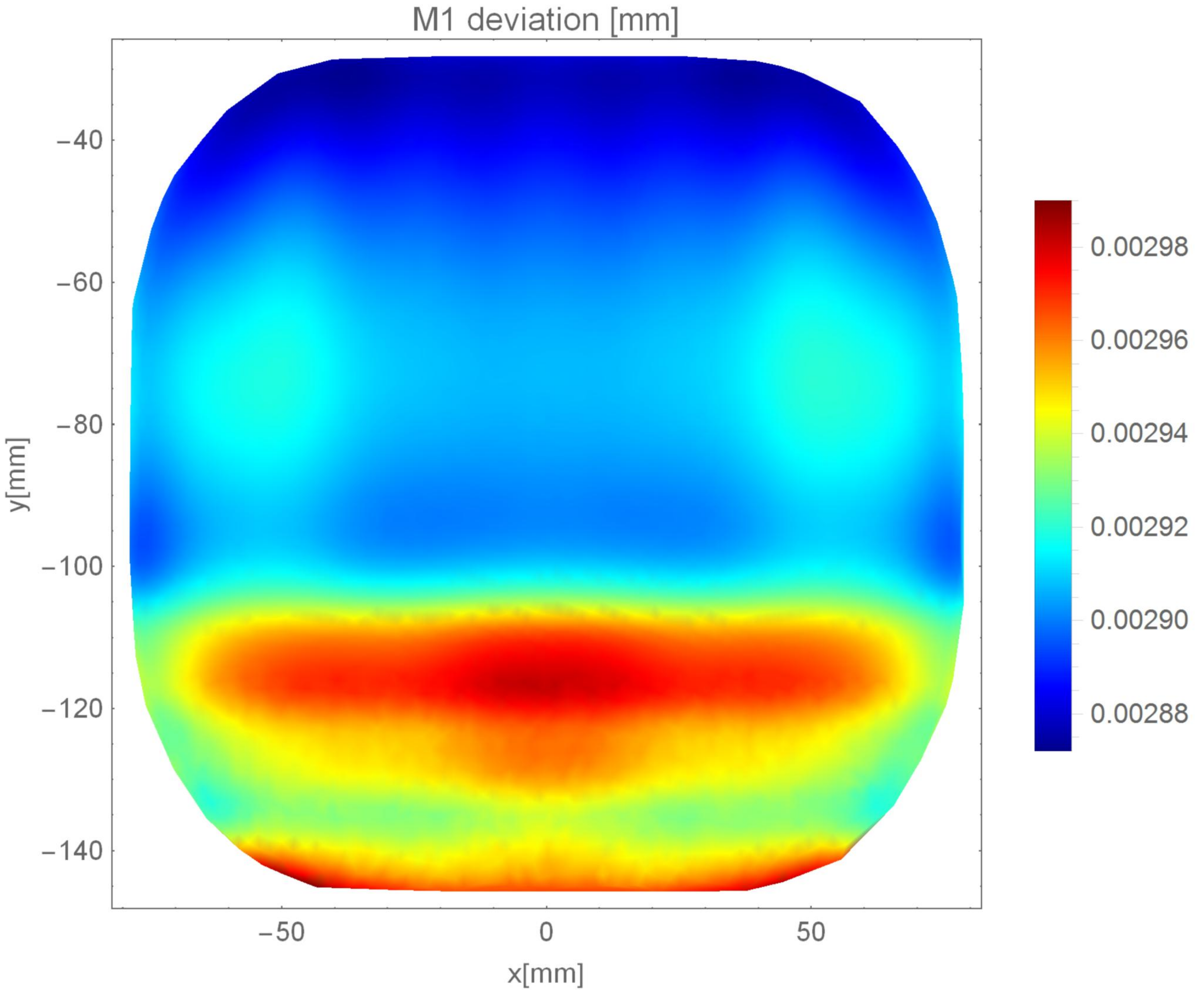}
\caption{\label{fig:m1fopexact}\resultcaption{1}{exactly compensated}{\monepvexcomp}{\monermsexcomp}{n}{}}
\end{minipage}\hfill
\begin{minipage}[t]{\figwidth\textwidth}
\includegraphics[width=\textwidth]{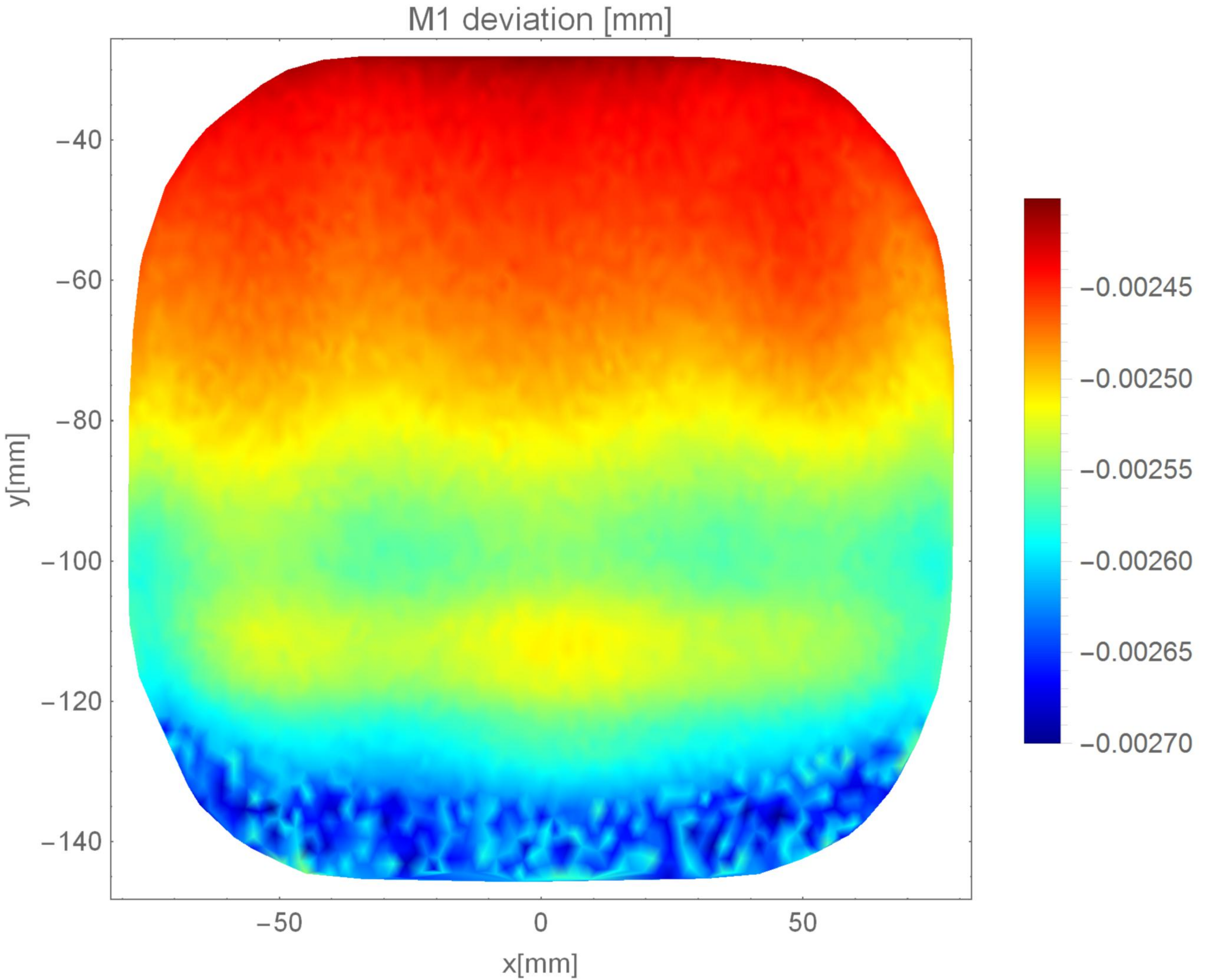}
\caption{\label{fig:m1fopcad}\resultcaption{1}{CAD compensated}{\monepvCADcomp}{\monermsCADcomp}{n}{PST/TIP/TLT: {\monepst}/$0^{\circ}$/{\monetlt},
after removal\\{p.-v./r.m.s.}: {\monefinalpvCADcomp} nm/{\monefinalrmsCADcomp} nm}}
\end{minipage}
\end{figure}
\begin{figure}
\begin{minipage}[t]{\figwidth\textwidth}
\includegraphics[width=\textwidth]{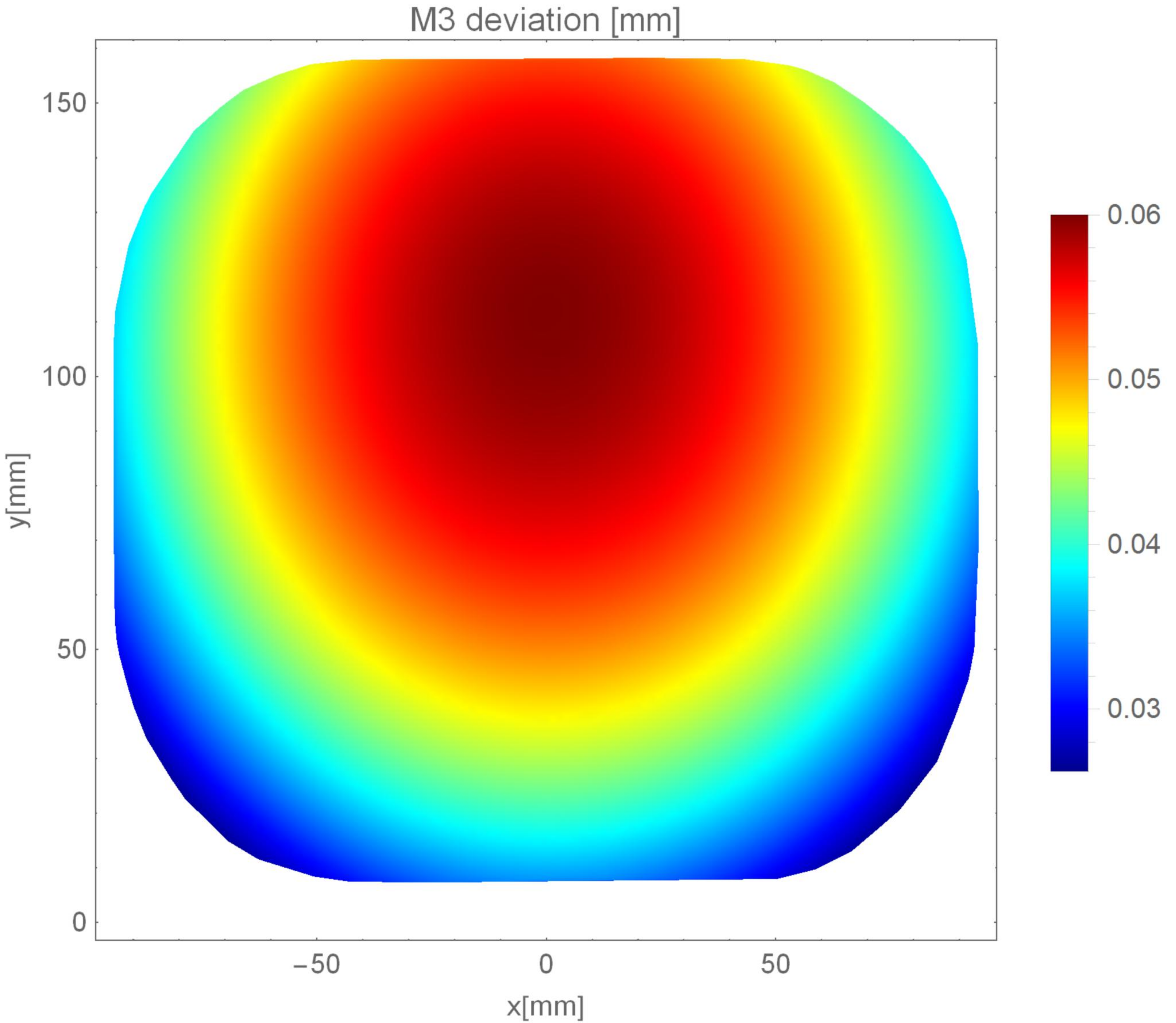}
\caption{\label{fig:m3nocomp}\resultcaption{3}{uncompensated}{\mthreepvnocomp}{\mthreermsnocomp}{$\mu$}{}}
\end{minipage}\hfill
\begin{minipage}[t]{\figwidth\textwidth}
\includegraphics[width=\textwidth]{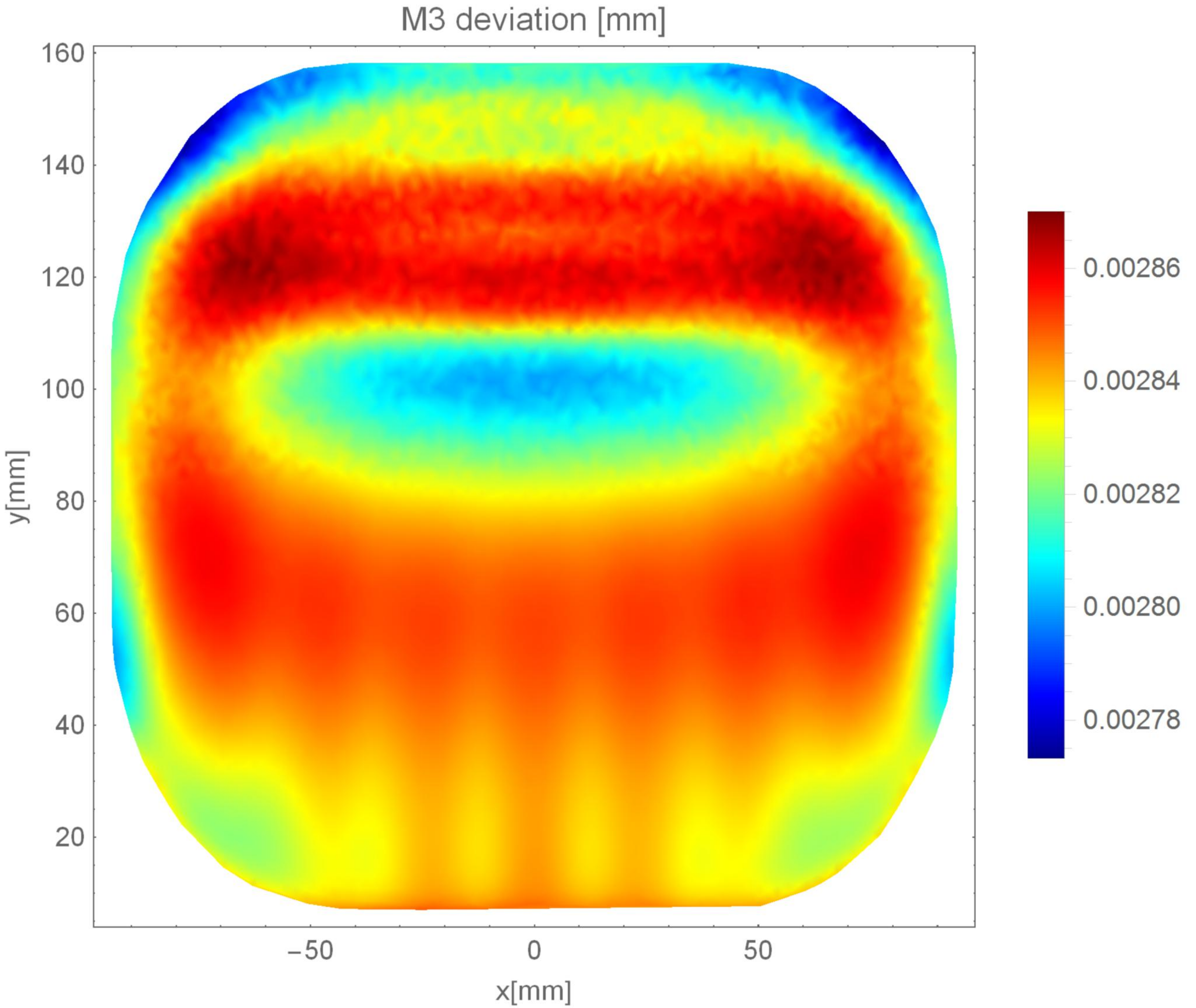}
\caption{\label{fig:m3fopexact}\resultcaption{3}{exactly compensated}{\mthreepvexcomp}{\mthreermsexcomp}{n}{}}
\end{minipage}\hfill
\begin{minipage}[t]{\figwidth\textwidth}
\includegraphics[width=\textwidth]{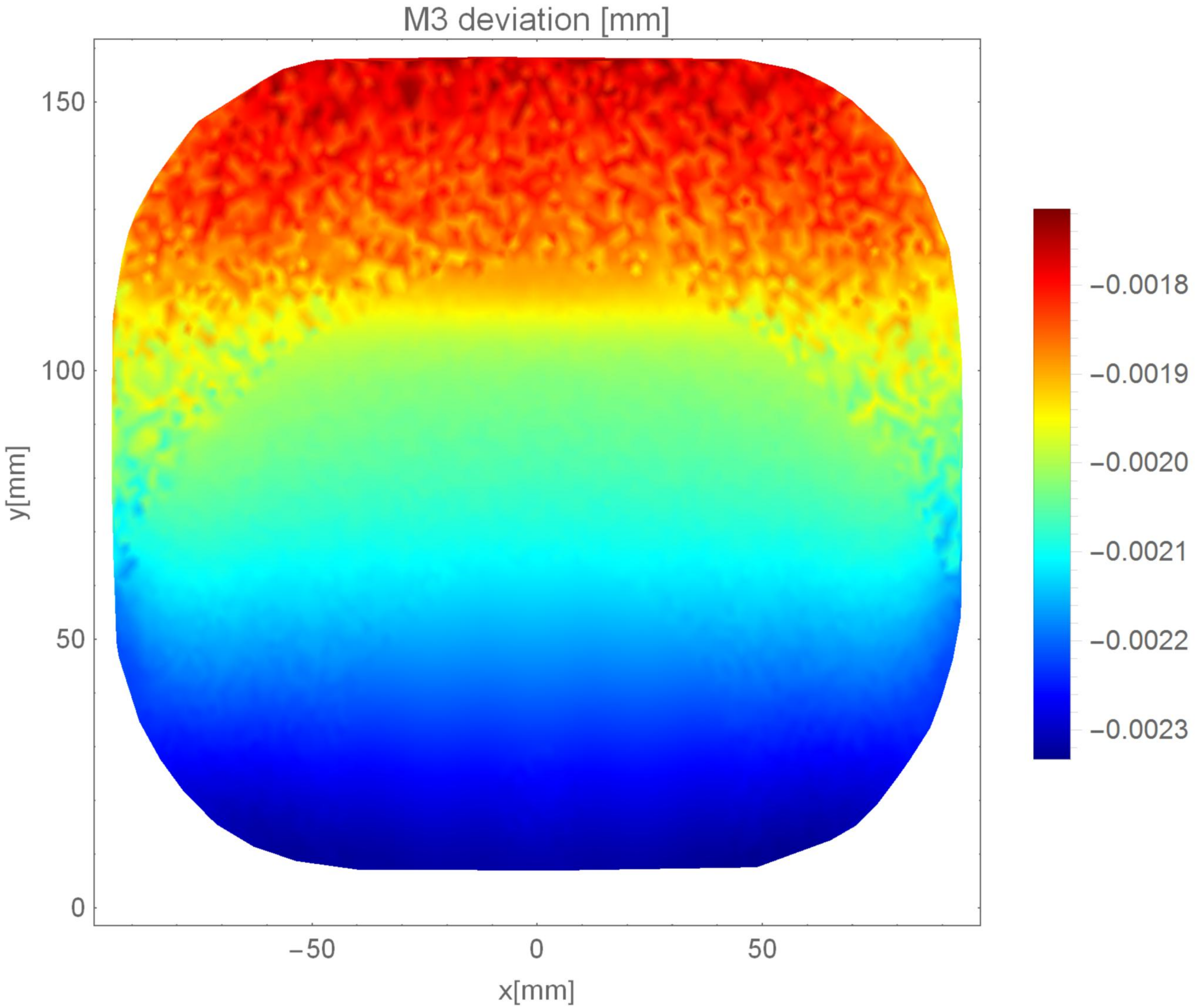}
\caption{\label{fig:m3fopcad}\resultcaption{3}{CAD compensated}{\mthreepvCADcomp}{\mthreermsCADcomp}{n}{PST/TIP/TLT: {\mthreepst}/$0^{\circ}$/{\mthreetlt}, after removal\\{p.-v./r.m.s.}: {\mthreefinalpvCADcomp} nm/{\mthreefinalrmsCADcomp} nm}}
\end{minipage}
\end{figure}

The FEA results show that the thermal deformation due to the cooling of a highly asymmetric {\material} mirror substrate 
with a NiP layer from room temperature down to {\cooltemp} may be compensated within two approaches. Those are also able to achieve 
a low figure error for the CAD compensated and for the exactly compensated model. The usual form deviation for the thermally matched
material combination without any compensation in this temperature range is of several microns for M1 and M3, respectively. 
Therefore, the compensation approach may be used to increase 
performance significantly for freeform mirrors that are not manufactured at operation temperature, for non-thermally matched mirror substrates with
their respective polishing coatings, or even for substrate materials that are both thermally matched.

\section{Summary and Outlook}\label{sec:summary}

The magnitude of the compensation in case of thermal loads is of an order accessible by diamond turning
or other ultra-precise manufacturing steps. But even if the temperature loads are small, the order of magnitude is due to its mainly low spatial
frequency content still accessible by MRF figure correction. It is emphasized that cryogenic thermal loads discussed throughout the present article
are an example for well-defined pre-compensable static loads. Although the compensation approach is usable for substrates, which
operate in, e.g., a high-temperature environment. The general formula \eqref{eq:compensation} does not make assumptions about the type of
deformation.

In the future the experimental verification of the derived compensation approach has to be carried out. The present article
contains theoretical considerations and the verification of the compensation was only done by FEA. Next step is to manufacture
a compensated substrate and perform the appropriate figure measurements by interferometric techniques under cryogenic conditions. 

\section{Acknowledgements}
The authors thank Matthias~Beier, Andreas~Gebhardt, Thomas~Peschel, and Stefan~Risse for very useful and enlighting discussions.
Further they thank Stephanie~Hesse-Ertelt and Aoife~Brady for helpful corrections.
This work is partly funded 
by Federal Ministry of Education and 
Research (BMBF) within framework ``Unternehmen Region -- Innovative Regional Growth Core'' 
grant number 03WKCK1B.



\input{uv_paper_arxiv_refs.bbl}

\end{document}

%% file: uv_paper_arxiv_refs.bbl
\providecommand{\href}[2]{#2}\begingroup\raggedright\endgroup